\documentclass[reprint,a4paper,australian,amsmath,aps,prd,twocolumn,superscriptaddress,preprintnumbers,noshowpacs]{revtex4-1}
\usepackage{times}
\usepackage{graphicx}
\usepackage{amssymb}
\usepackage{amsmath}
\usepackage{mathrsfs}
\usepackage[utf8]{inputenc}
\usepackage{xcolor}
\usepackage{dcolumn}
\usepackage{bm}
\usepackage{yhmath}
\usepackage{hyperref}
\usepackage{booktabs}
\usepackage{subfigure}
\usepackage{color}
\usepackage{multirow}
\usepackage{dcolumn}
\usepackage[normalem]{ulem}

\newcolumntype{C}[1]{>{\centering\arraybackslash}p{#1}}

\newcommand{\be}{\begin{equation}}
\newcommand{\ee}{\end{equation}}

\newcommand{\CSSM}{Special Research Centre for the Subatomic Structure
  of Matter (CSSM),\\Department of Physics, University of
  Adelaide, Adelaide, South Australia 5005, Australia}

\newcommand{\CoEPP}{ARC Centre of Excellence for Particle Physics at
  the Terascale (CoEPP),\\Department of Physics, University
  of Adelaide, Adelaide, South Australia 5005, Australia}

\newcommand{\Lanzhou}{School of Physical Science and Technology, Lanzhou University, Lanzhou 730000, China }

\begin{document}
\preprint{ADP-17-13/T1019}
\title{
{Structure of the Roper Resonance from Lattice QCD Constraints}
}
\author{Jia-jun Wu}
\author{Derek B. Leinweber}
\affiliation{\CSSM}
\author{Zhan-wei Liu}
\affiliation{\CSSM}
\affiliation{\Lanzhou}
\author{Anthony W. Thomas}
\affiliation{\CSSM}
\affiliation{\CoEPP}
\pacs{
{12.39.Fe}{ Chiral Lagrangians}
{12.38.Gc}{ Lattice QCD calculations}
}

\begin{abstract}
Two different effective field theory descriptions of the pion-nucleon
scattering data are constructed to describe the region of the Roper
resonance.
In one, the resonance is the result of strong rescattering between
coupled meson-baryon channels, while in the other the resonance has a
large bare-baryon (or quark-model like) component.  The predictions of
these two scenarios are compared with the latest lattice QCD
simulation results in this channel.  We find that the second scenario
is not consistent with lattice QCD results, whereas the first agrees
with those constraints.  In that preferred scenario, the mass of the
quark-model like state is approximately 2 GeV, with the
infinite-volume Roper resonance best described as a resonance
generated dynamically through strongly coupled meson-baryon channels.
\end{abstract}

\maketitle

\section{Introduction}
\label{sect:intro}

Since the discovery of the Roper resonance in 1964~\cite{Roper:1964zza}, its peculiar properties
have challenged our understanding of the quark structure of hadrons and ultimately of quantum
chromodynamics (QCD) itself \cite{Isgur:1978wd, Aznauryan:2008pe, Joo:2005gs, Weber:1989fv,
  JuliaDiaz:2006av, BarquillaCano:2007yk, Golli:2007sa, Golli:2009uk, Meissner:1984un,
  Hajduk:1984ry, Krehl:1999km, Schutz:1998jx, Matsuyama:2006rp, Kamano:2010ud, Kamano:2013iva,
  Suzuki:2009nj, Hernandez:2002xk, Barnes:1982fj, Golowich:1982kx, Kisslinger:1995yw}.  With the
first negative-parity excitation of the nucleon, the $N^\star(1535)$, almost 600 MeV above the nucleon, 
expectations -- based upon the harmonic oscillator model which has enjoyed success in
treating hadron spectroscopy -- suggest that the first positive-parity excited state should occur
around 2 GeV.  Yet, empirically one finds the first positive-parity, spin-1/2 Roper
resonance of the nucleon
to have a mass of just 1.45 GeV, {\it below} the $N^\star(1535)$ ~\cite{Olive:2016xmw}!

To make matters worse, the first negative-parity excitation of a strangeness -1 baryon, the famous
$\Lambda(1405)$, is lower in mass than both of these non-strange excited states of the nucleon
~\cite{Olive:2016xmw}. Fortunately, in this case there have recently been advances in our
understanding, via lattice QCD simulations of not only the mass of this state but the individual
valence quark contributions to its electromagnetic form factors~\cite{Hall:2014uca,Hall:2016kou}.
These simulations have been supported by analysis involving an effective Hamiltonian
\cite{Liu:2015ktc}, which allows a natural connection to be made between the results calculated on
a finite lattice volume and the infinite volume of the real world~\cite{Liu:2016wxq,
  Molina:2015uqp}. As a result of these studies, it is now clear that the $\Lambda(1405)$ is
essentially an anti-kaon nucleon bound state with very little content corresponding to the sort of
three-quark state anticipated in a typical quark model~\cite{Hall:2016kou}.

In this article we use similar techniques to investigate the nature of the Roper
resonance.  Our calculations are founded on Hamiltonian effective field theory (HEFT), an extension
of chiral perturbation theory that incorporates the L\"uscher relation
\cite{Luscher:1986pf,Luscher:1990ux,Luscher:1991cf} connecting the energy levels observed in finite
volume to the scattering phase shifts \cite{Wu:2014vma}.  In the power-counting regime, HEFT
reproduces the expansion of chiral perturbation theory for ground state phenomena
\cite{Young:2002ib}.

The results presented herein are the first to incorporate a basis state that can be associated with
a quark-model state for the Roper, where radial excitations of constituent quarks describe the
internal structure of the Roper.  This is an important development that admits, for example,
three-quark descriptions of nucleon-Roper transition form factors in the large momentum transfer
regime \cite{Wilson:2011aa} where mesonic dressings are suppressed.

The outline of this article is as follows.  We first introduce the coupled-channel scattering
formalism~\cite{Liu:2016uzk} in Sec.~\ref{framework}.  Experimental scattering data in the region
of the Roper resonance is analyzed in Sec.~\ref{data} where two different descriptions of the data
are obtained in the coupled-channel formalism.  In the first fit there is no significant three
quark coupling, while in the second alternative fit to the data there is.  These models produce rather
different behaviour in the unobserved $\pi\Delta$ and $\sigma N$ channels and cannot be
distinguished by experiment.  We then use the same effective field theory on a finite volume in
Sec.~\ref{lattice} to compute the spectrum one would expect to find in lattice QCD.  Only the first
description of the experimental data is consistent with recent lattice simulations, indicating that
the Roper resonance is generated dynamically through the rescattering of coupled meson-baryon
channels.  

This discovery leads to a new contemporary role for constituent quark models in describing the
low-lying baryon spectrum.  Section~\ref{role} presents this role drawing on recent developments in
the understanding of the odd-parity $N$ and $\Lambda$ spectra.  The three-quark radial excitation
of the nucleon anticipated in traditional quark models appears to lie closer to 2
GeV, which as explained earlier, is in accord with the excitation energy of the observed
$N^\star(1535)$.  Finally, Sec.~\ref{sec:con} summarizes our conclusions and suggests directions
for future research.

\section{Theoretical framework}
\label{framework}

In order to model the scattering data in the region of the Roper resonance and
describe the observed inelasticity, we include three  coupled channels,
$\pi N$, $\pi \Delta$ and $\sigma N$. In the rest frame, the Hamiltonian
has the following form
\begin{eqnarray}
H = H_0 + H_I \, ,
\label{eq:h}
\end{eqnarray}
where the non-interacting Hamiltonian is
\begin{eqnarray}
H_0 &=&\sum_{B_0} |B_0\rangle \, m^{0}_{B} \, \langle B_0|+ \sum_{\alpha}\int d^3\vec{k}\nonumber\\
&&  |\alpha(\vec{k})\rangle \, \left [ \omega_{\alpha_1}(\vec{k})
+ \omega_{\alpha_2}(\vec{k})\, \right ] \langle\alpha(\vec{k})| \, .
\label{eq:h0}
\end{eqnarray}
Here $B_0$ denotes a bare baryon with mass $ m^{0}_{B}$, which may be thought of
as a quark model state and $\alpha_1$ ($\alpha_2$) indicates the meson (baryon) 
state which constitutes channel $\alpha$, with $\omega_{\alpha_i}(\vec{k})
= \sqrt{m^2_{\alpha_i}+\vec{k}^2}$.

The energy independent interaction Hamiltonian includes two parts, $H_I = g + v$,
where $g$ describes the vertex interaction between the bare particle
and the two-particle channels $\alpha$
\begin{eqnarray}
g &=& \sum_{\alpha,\, B_0} \int d^3\vec{k} \left \{  \,
|\alpha(\vec{k})\rangle \, G^\dagger_{\alpha, B_0}(k)\, \langle B_0|+h.c.
\right \}\, ,
\label{eq:int-g1}
\end{eqnarray}
while the direct two-to-two particle interaction is defined by
\begin{eqnarray}
v = \sum_{\alpha,\beta} \int d^3\vec{k}\; d^3\vec{k}'\,
|\alpha(\vec{k})\rangle\,  V^{S}_{\alpha,\beta}(k,k')\, \langle
\beta(\vec{k}')| \, .
\label{eq:int-v}
\end{eqnarray}
For the vertex interaction between the bare baryon and the two-particle
channels we choose:
\begin{eqnarray}
G_{\alpha, B_0}^2(k)&=&\frac{g_{B_0\alpha}^2}{4\pi^2}\left (\frac{k}{f}\right )^{2l_\alpha} \frac{ u_{\alpha}^2(k)}{\omega_{\alpha_1}(k)} \, ,
\end{eqnarray}
where the pion decay constant $f=92.4$ MeV and $l_\alpha$ is the orbital angular momentum in
channel $\alpha$. Here, since we are concerned with the Roper resonance, with isospin, angular
momentum and parity, $\mathbf{I(J^P)=\frac12(\frac12^+)}$, $l$ is 1 for $\pi N$ and $\pi \Delta$,
while it is 0 for $\sigma N$.  The regulating form factor, $u_\alpha(k)$, takes the exponential
form
%
$u_\alpha(k)=\exp\left ( -{k^2}/{\Lambda_\alpha^2} \right ) ,$
%
where $\Lambda_\alpha$ is the regularization scale.
For the five direct two-to-two particle interactions we introduce separable
potentials 
\begin{eqnarray}
V_{\alpha, \beta}^S(k,k')&=&g^S_{\alpha, \beta}\frac{\bar G_{\alpha}(k)}{\sqrt{\omega_{\alpha_1}(k)}}
      \frac{\bar G_{\beta}(k')}{\sqrt{\omega_{\beta_1}(k')}}\, ,
\label{eqVspiN}
      \end{eqnarray}
where $\bar G_{\alpha}(k)=G_{\alpha, B_0}(k)/g_{B_0 \alpha}$.
The $T$-matrices for two particle scattering are obtained by solving
a three-dimensional reduction of the coupled-channel Bethe-Salpeter
equations for each partial wave
\begin{eqnarray}
t_{\alpha, \beta}(k,k';E)&=&V_{\alpha, \beta}(k,k';E)+\sum_\gamma \int
q^2 dq \times \\
&&\hspace{-2.5cm} V_{\alpha, \gamma}(k,q;E)\,
\frac{1}{E-\omega_{\gamma_1}(q)-\omega_{\gamma_2}(q)+i \epsilon}\,  t_{\gamma, \beta}(q,k';E)
\, . \nonumber
\end{eqnarray}
The coupled-channel potential is readily calculated from the interaction Hamiltonian
\begin{eqnarray}
V_{\alpha, \beta}(k,k') &=& \sum_{B_0}\,
\frac{G^\dag_{\alpha, B_0}(k)\, G_{\beta, B_0}(k')}{E-m_B^0} +V^S_{\alpha,\beta}(k,k') \, ,
\label{eq:lseq-2}
\end{eqnarray}
with the normalization $\langle \alpha(\vec{k})\, |\, \beta(\vec{k}^{\,\,'})\rangle =
\delta_{\alpha,\beta}\, \delta (\vec{k}-\vec{k}^{\,\,'})$.
The pole position of any bound state or resonance is obtained by searching for the
poles of the $T$-matrix in the complex plane.
\begin{figure}[tbp]
\begin{center}
\includegraphics[width=1.0\columnwidth]{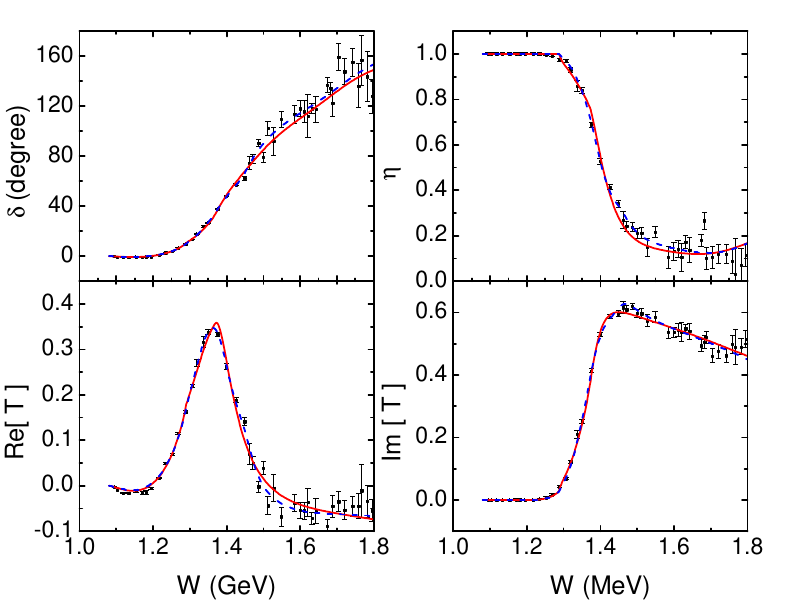}
\caption{
%
The fitted phase shift $\delta$, inelasticity $\eta$ and $T$-matrix
for the $\pi N \to \pi N$ reaction. Red-solid and blue-dashed lines
are calculated from scenarios I and II, respectively.
%
}\label{fg:phase}
\end{center}
\end{figure}

In order to compare the predictions of this infinite-volume model with the results
of lattice QCD simulations, it is necessary to rewrite the problem on a finite volume.  The details
of this procedure are described in Refs.~\cite{Liu:2015ktc, Liu:2016wxq, Liu:2016uzk, Hall:2013qba,
Wu:2014vma, Wu:2016ixr}.  By solving for the eigenstates of the HEFT 
one obtains energy levels and eigenstates which can be compared with the energies
and interpolating fields in the lattice QCD simulations. 

We can also extend the formalism to unphysical
pion masses.  Using $m_\pi^2$ as a measure of the light quark masses, we consider the variation of
the bare mass and $\sigma$-meson mass as
\begin{eqnarray}
m_B^0(m_\pi^2)&=&m_B^0|_{\rm phy}+\alpha_B^0\, (m_\pi^2-m_\pi^2|_{\rm phy})\, , \label{Eq:alpha} \\
m_\sigma^2(m_\pi^2)&=&m_\sigma^2|_{\rm phy}+\alpha_\sigma^0\, (m_\pi^2-m_\pi^2|_{\rm phy}) \, ,
\end{eqnarray}
where the slope parameter $\alpha_B^0$ is constrained by lattice QCD data from the CSSM.  In the
large quark mass regime, where constituent quark degrees of freedom become relevant, one
expects~\cite{Cloet:2002eg} $\alpha_\sigma^0 = (2/3) \alpha_N^0$.
The nucleon and Delta masses away from the physical point are obtained via linear interpolation
between the lattice QCD results.

The incorporation of couplings between the three 
scattering channels requires 6 parameters and coupling of these channels to a bare basis state
requires another three.  Regularization 
admits four fit parameters.  This description is typical of contemporary analyses of resonance
phenomena~\cite{JuliaDiaz:2007kz,Arndt:2006bf,Ronchen:2012eg} and ensures accuracy in connecting
experiment to the finite-volume spectrum.

\section{Experimental Data Analysis}
\label{data}

By fitting the experimental data for $\pi N$ scattering from 1200 MeV to 1800 MeV,
we found two different parameter sets which appear equally acceptable in describing existing
data.  The phase shifts and inelasticities for the $\pi N \to \pi N$ channel are shown
for these two scenarios in Fig.~\ref{fg:phase}.  The parameter sets are described in
Table~\ref{tab:parameter}.
The $\chi^2$ for scenarios I and II are 241 and 135 respectively.  The larger $\chi^2$ for scenario
I has its origin in just 6 points at the opening of the $\pi \Delta$ channel and can be attributed to
the zero-width approximation for the $\Delta$.  In the context of the benchmark result
\cite{GWdataAnalysisCentre} of $\chi^2 = 236$, both fits may be regarded as an acceptable
characterization of the experimental data.

In Scenario I, the coupling of $\pi N \to \pi N$ is enhanced while the coupling of the bare state
to $\pi N$ and $\sigma N$ is suppressed relative to Scenario II.  This presents two different
pictures of the Roper. In Scenario I, the Roper is a resonance generated by strong rescattering in
the meson-baryon channels.  In Scenario II the rescattering is weaker and the observed resonance is
dominated by coupling to an underlying bare, or quark-model like, state.  It is this latter
scenario that the community has anticipated for the Roper since the advent of the constituent quark
model \cite{Isgur:1978wd}.

\begin{figure}[tbp]
\begin{center}
\includegraphics[width=0.5\textwidth]{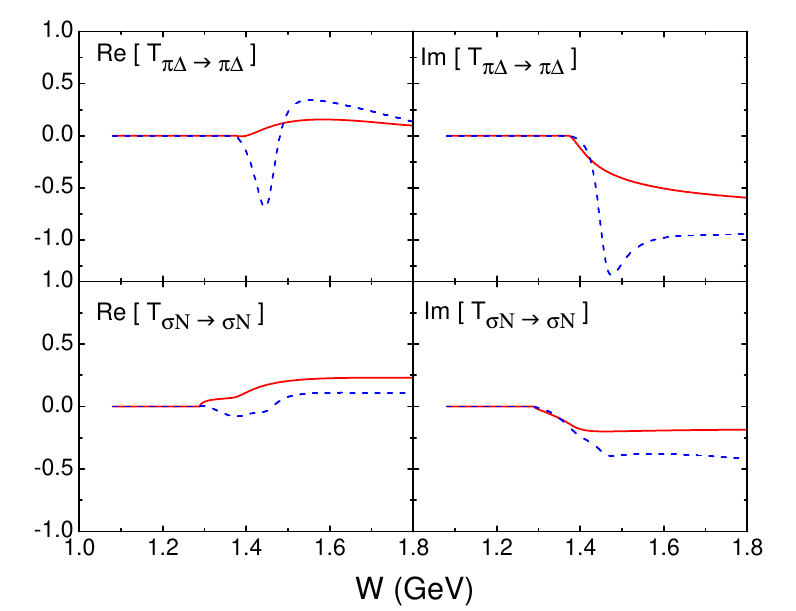}
\caption{
%
The $T$-matrices for $\pi \Delta \to \pi \Delta$ and $\sigma N \to \sigma N$ corresponding to the
two scenarios described in the text.  The two cases are encoded as in Fig.~\ref{fg:phase} where
red-solid and blue-dashed lines are calculated from scenarios I and II, respectively.
}\label{fg:Tother}
\end{center}
\end{figure}

\begin{table}[bp]
\caption{Fit parameters constrained by $\pi N$ scattering data and the resultant pole positions in
  the two scenarios described in the text.
  The pole position in the different Riemann sheets is also indicated for each channel ($\pi N$,
  $\pi \Delta$, $\sigma N$).  The unphysical sheet is denoted "u" and the physical sheet is denoted
  "p", as defined in Ref.~\cite{Suzuki:2008rp, Doring:2009yv}.
}
\label{tab:parameter}
\begin{ruledtabular}
\begin{tabular}{ccc}
\noalign{\smallskip}
Parameter                                           & I                                  & II   \\
\noalign{\smallskip}
\hline
\noalign{\smallskip}
$ g^S_{\pi N, \pi N} $                         & $1.156$                      & $0.634$                      \\
$ g^S_{\pi N, \pi \Delta} $                  & $-0.662$                    & $-0.378$                      \\
$ g^S_{\pi N, \sigma N}  $                  & $-0.415$                    & $-1.738$                      \\
$ g^S_{\pi \Delta, \pi \Delta} $           & $-0.438$                    & $-0.581$                      \\
$ g^S_{\pi \Delta, \sigma N}  $           & $1.332$                     & $0.964$                      \\
$ g^S_{\sigma N, \sigma N}   $           & $10.000$                   & $10.000$                      \\
$ m_B^0/{\rm GeV} $                          & $2.000$                     & $1.7000$   \\
$ g_{B_0\pi N}  $                                 & $0.268$                     & $0.954$          \\
$ g_{B_0\pi\Delta} $                            & $1.544$                     & $-0.118$         \\
$ g_{B_0\sigma N}  $                           & $ - $                           & $-2.892$         \\
$ \Lambda_{\pi N}/{\rm GeV} $           & $0.5953$                   & $0.6302$                      \\
$ \Lambda_{\pi \Delta}/{\rm GeV} $   & $1.5000$                    & $1.4318$                       \\
$ \Lambda_{\sigma N}/{\rm GeV}  $   & $1.5000$                    & $1.4533$                        \\
\noalign{\medskip}
Pole  (MeV) (uuu)                                & $2012.28 - 42.09\,i$            & $1355.57 - 70.81\,i$               \\
Pole  (MeV) (upu)                                & $1392.92 - 167.13\,i$            & $1362.33 - 100.53\,i$               \\
\end{tabular}
\end{ruledtabular}
\end{table}

While both scenarios describe the present experimental data, they make unique predictions in the
coupled channels $\pi \Delta \to \pi \Delta$ and $\sigma N \to \sigma
N$ as illustrated in Fig.~\ref{fg:Tother}.
Measurements of these scattering amplitudes would enable us to distinguish between the scenarios.

\section{Lattice QCD Constraints}
\label{lattice}

In the absence of the relevant experimental data, we now turn to the results provided by lattice
QCD simulations, focusing on the recent work of Lang {\it et al.}~\cite{Lang:2016hnn} and the
CSSM~\cite{Liu:2016uzk}.
HEFT predictions in the finite volume of the lattice are constrained at the physical quark masses
by the fit to experimental data.  To extend the predictions to other quark masses, we introduce
one new parameter, $\alpha_B^0$ of Eq.~(\ref{Eq:alpha}), providing a linear quark-mass dependence
in the bare mass.
Noting that the lattice simulation at the lightest quark mass involves a pion mass of 156
MeV, our key results are insensitive to this extension to larger pion masses.

Lang {\it et al.}'s work \cite{Lang:2016hnn} is particularly interesting as it incorporates $\pi N$
and $\sigma N$ five-quark non-local interpolating fields with the momenta of each hadron projected
to provide excellent overlap with the low-lying scattering states of the spectrum.
These results are particularly important in discriminating between Scenarios I and II.

These non-local interpolators are complemented by standard local
three-quark interpolating fields, which have proved to favour localized states and miss the
non-local scattering states
\cite{Mahbub:2010rm,Mahbub:2012ri,Mahbub:2013ala,Mahbub:2013bba,Kiratidis:2015vpa,Kiratidis:2016hda}, 
making the lattice spectrum incomplete.
In this case there may be concern that the lattice energy levels extracted might be systematically
contaminated from the missed levels.  The CSSM collaboration has explored this in detail
\cite{Mahbub:2013bba,Kiratidis:2015vpa,Kiratidis:2016hda} and has developed methods that ensure
these systematic errors are suppressed through Euclidean time evolution.  Their use of a
single-state Ansatz and a full covariance-matrix calculation of the $\chi^2/{\rm dof}$ with a
conservative cutoff of $\chi^2/{\rm dof} < 1.2$ ensures that any remaining contamination is contained
within the error bars reported \cite{Kiratidis:2016hda}.  While the resultant uncertainties overlap
with several states in the HEFT spectrum, we'll see that it is the absence of localized states
below 1.9 GeV \cite{Dudek:2012ag,Kiratidis:2016hda} that facilitates a discrimination of Scenarios I and II.
\begin{figure}[t]
\begin{center}
\includegraphics[width=1.0\columnwidth]{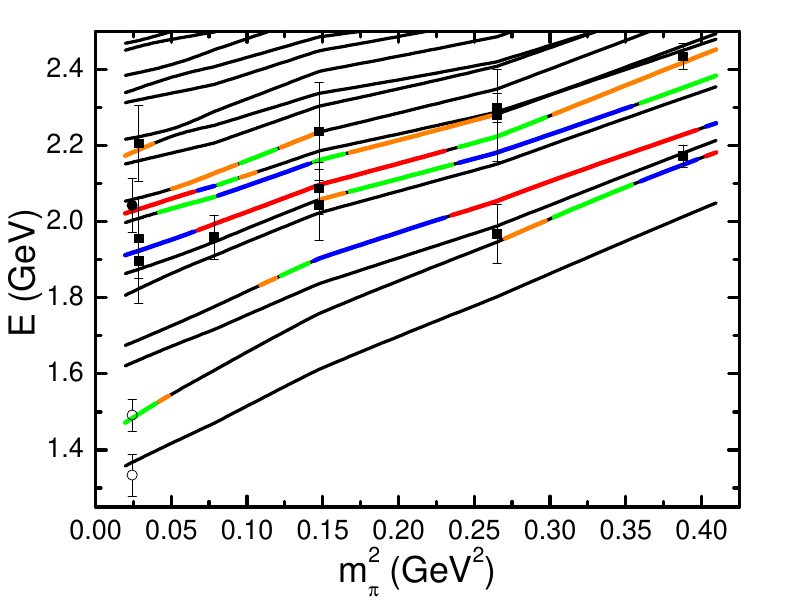}
\caption{
The finite volume spectrum of Scenario I with a bare mass of 2.0 GeV.  The CSSM results
\cite{Liu:2016uzk} are indicated by square symbols and circles denote the more recent results from
Lang {\it et al.}\ \cite{Lang:2016hnn}.  Solid symbols indicate states dominated by local
three-quark operators and open symbols indicate states dominated by non-local momentum-projected
five-quark operators.  The colours red, blue, green and orange are used to indicate the relative
contributions of the bare-baryon basis state in the eigenstate, with red being the largest
contribution.
}\label{fg:specI}
\end{center}
%
\begin{center}
\includegraphics[width=1.0\columnwidth]{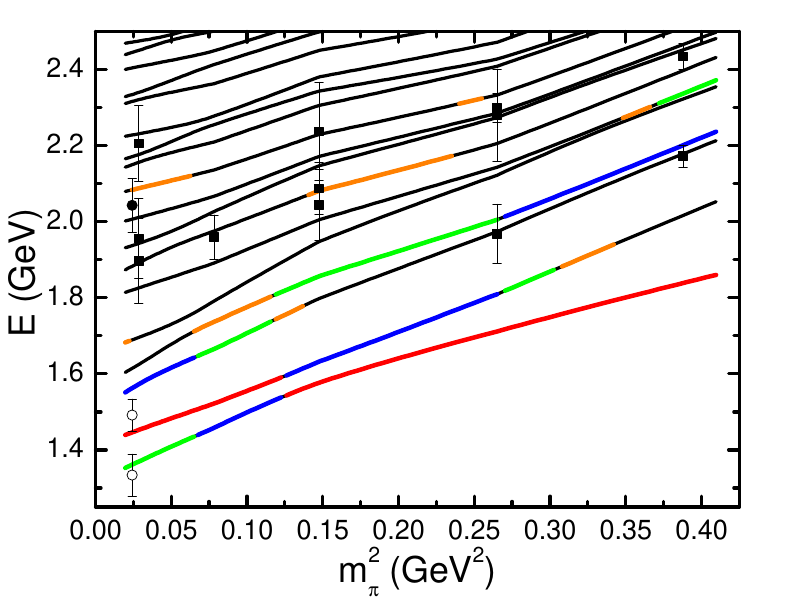}
\caption{
The finite volume spectrum corresponding to Scenario II having a
bare mass of 1.7 GeV. Results are illustrated as described in
Fig.~\ref{fg:specI}.
}\label{fg:specII}
\end{center}
\end{figure}
\begin{figure}[tbh]
\begin{center}
\includegraphics[width=1.0\columnwidth]{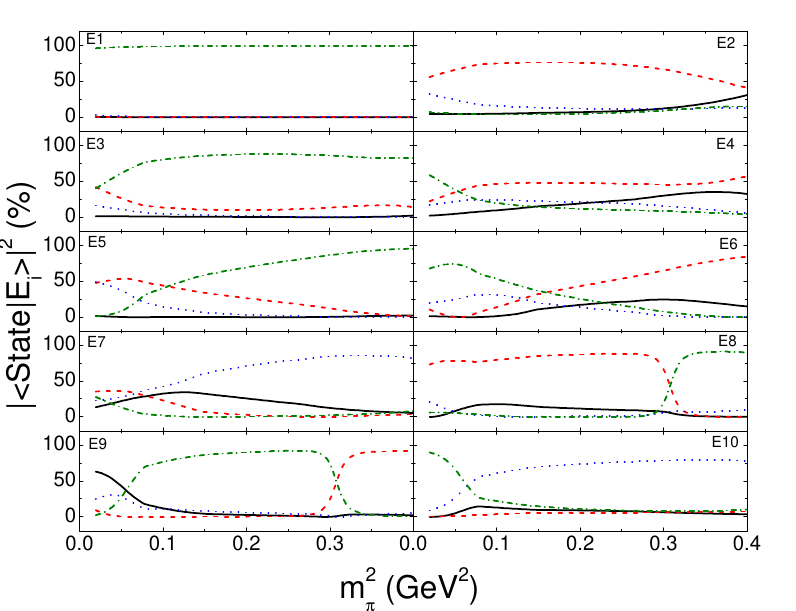}
\caption{
The pion-mass evolution of the Hamiltonian eigenvector components for Scenario I having a bare mass
of 2.0 GeV. The black solid line indicates the fraction of the bare-state, $|\langle
m_0|E_i\rangle|^2$.  The red dashed, blue dotted and green dashed-dotted lines show the $\pi N$,
$\pi \Delta$ and $\sigma N$ ``State'' contributions to the energy eigenstate $| E_i\rangle$
respectively, summed over all discrete momenta, $\sum_{\vec k} |\langle \alpha(\vec k) |
E_i\rangle|^2$.  The labels ``Ei'' indicate the i'th energy level.
}\label{fg:waveallI}
\end{center}
\end{figure}

In reporting the lattice QCD results in Figs.~\ref{fg:specI} and \ref{fg:specII} we have used
solid (open) symbols to indicate states dominated by local (non-local) interpolating fields.
While the operator overlaps are unrenormalized matrix elements and therefore scale
  dependent, the qualitative aspects of this information can be used to gain insight into the
  composition of the states in addition to the standard analysis of the spectrum.
Just as insight into the composition of the lattice QCD states can be obtained from the
eigenvectors of the lattice correlation matrix used to excite the states, HEFT also
provides insight into their composition via the superposition of basis states in each eigenvector.

B\"ar \cite{Bar:2017kxh} has shown that the overlap of the non-interacting non-local $\pi N$ channel with
standard local three-quark interpolating fields is suppressed by three orders of magnitude relative
to the ground state nucleon on a 3 fm lattice.  Thus the lattice QCD states that are excited by
local three-quark operators correspond to HEFT states having a significant bare-basis-state
component in their eigenvector.  This information is indicated through the colour coding of the
HEFT spectra illustrated in Figs.~\ref{fg:specI} and \ref{fg:specII} for Scenarios I and II
respectively. The red, blue, green and orange lines indicate the states having the first, second, third and
fourth largest bare-baryon basis-state contributions, respectively.

For Scenario I of Fig.~\ref{fg:specI}, all of the lattice states dominated by local three-quark
interpolating fields can be associated with a colored line.  Similarly, all of the Hamiltonian
states having the largest bare basis-state component, indicated in red in Fig.~\ref{fg:specI}, have
a nearby lattice QCD result.  We quantify this agreement through a $\chi^2$ measure associating
each three-quark dominated (solid) lattice point with a three-quark dominated HEFT state.  Open
symbols are associated with the nearest HEFT level. For Scenario I the minimum $\chi^2/{\rm dof} =
16.5/(15-1) = 1.18$.

On the other hand, Scenario II of Fig.~\ref{fg:specII} displays little correspondence to the
lattice QCD results.  Scenario II predicts a low-lying state with a large bare basis-state
component of approximately 50\%, approaching that for the ground state.  Such a state would be easy
to excite in lattice QCD with local three-quark operators.  However this three-quark dominated
state is not seen in the simulations.  For this scenario, the $\chi^2$ measure provides
$\chi^2/{\rm dof} = 635.7/(15-1) = 45.4$.

We also consider a simple $\chi^2$ measure where only the most bare-state dominated HEFT state and
the closest three-quark-interpolator dominated lattice QCD point are considered at each mass.  In
Scenario I the $\chi^2/{\rm dof} = 5.78/(6-1) = 1.16$, whereas in Scenario II the $\chi^2/{\rm dof}
= 338/(6-1) = 68$.

Focusing now on the lattice QCD results from Lang {\it et al.}\ \cite{Lang:2016hnn} at near physical
quark masses, we see that both the energy levels and their composition are correctly described in
Scenario I of Fig.~\ref{fg:specI}.
The lowest-lying lattice-QCD state appears with the introduction of their momentum-projected
$\sigma N$ interpolator.  The second state disappears if their momentum-projected $\pi N$
interpolator is omitted \cite{Lang:2016hnn}.
This composition agrees with the composition predicted by HEFT in Scenario I reported in
Fig.~\ref{fg:waveallI}.
Near the physical mass the first eigenstate is dominated by $\sigma N$ basis states.  The second
state is dominated by the $\pi N$ channel complemented by some mixing with $\pi \Delta$.

While the HEFT of Scenario I correctly describes the composition of these states, Scenario II
describes the lattice $\pi N$ state as a three-quark dominated state with 50\% of the composition
in the local three-quark state.  Yet three-quark interpolators do not see this state.  This is an
important discrepancy which again excludes Scenario II as an acceptable description of the nucleon
spectrum.

It is only for the seventh, eighth and ninth eigenstates of Scenario I that we
find a significant bare basis-state contribution in Fig.~\ref{fg:waveallI} and
this is precisely where the lattice QCD states excited by local three-quark operators reside.
The ninth state has an extremely large bare-state component exceeding 50\% and both the CSSM and
Lang {\it et al.}\  observe three-quark dominated lattice QCD eigenstates within
one sigma of this state, lending further credence to Scenario I as the correct
description of the nucleon spectrum.

Finally, we examine how the eigenstates evolve as the pion mass increases.  In
Fig.~\ref{fg:waveallI} we see that in scenario I, the bare baryon content of the second and fourth
eigenstates increases towards the upper end of the pion-mass range.  Once again this is consistent
with the lattice simulations as this is where the CSSM finds lower mass states in the spectrum with
local three-quark operators.  Overall the Hamiltonian eigenvectors obtained within Scenario I
explain the lattice spectra very well.

In contrast, the lowest-mass eigenstates of Scenario II are dominated by the bare-baryon
basis state.  At near-physical quark masses, their composition is inconsistent with the results of
Lang {\it et al.}\ and at larger quark masses neither the CSSM nor the Cyprus collaboration
\cite{Alexandrou:2014mka} observed such low-lying states in their 3 fm lattice
results.

\section{Contemporary Role for the Quark Model}
\label{role}

Through a consolidation of the earlier Lattice QCD and HEFT approaches to the study of the
$N^*(1535)$ and $\Lambda^*(1405)$ resonances and the study of the $N^*(1440)$ resonance herein, a
new understanding of the nature of these resonances and the role of the quark model is emerging.
\begin{itemize}
\item[1.] The $N^*(1535)$ is dominated by a three-quark core and dressed by a meson cloud \cite{Liu:2015ktc}.  While
  the role of the meson cloud is enhanced, the structure of this state is qualitatively similar to
  ground state. \\

\item[2] The $\Lambda^*(1405)$ is predominantly a molecular $\bar{K} N$ bound state
  \cite{Hall:2014uca,Liu:2016wxq}.  Mixing with the $\pi \Sigma$ channel creates a resonance with
  a two-pole structure.
The excited state of the quark model lies higher at approximately 1.6 GeV.\\

\item[3] The $N^*(1440)$ resonance is best described as the result of strong rescattering between
  coupled meson-baryon channels with only a small component associated with a quark-model-like
  state.  This small but nontrivial contribution is indicated by the green curve in
  Fig.~\ref{fg:specI} through the second excitation reported by Lang {\it et al.}
  \cite{Lang:2016hnn}.
  The first radially excited nucleon of the quark model has a mass of approximately 2 GeV.
\end{itemize}

\begin{figure}[tpb]
\includegraphics[width=1.06\columnwidth]{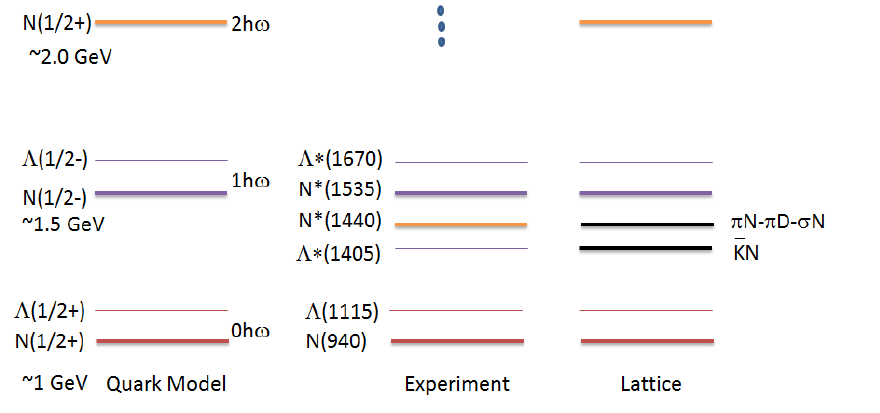}
\caption{
The low-lying $N$ and $\Lambda$ spectra of the simple quark model described herein are presented in
the context of experimental data and the results of Lattice QCD plus HEFT analyses as described in
the text.
}\label{fg:newp}
\end{figure}

These conclusions provide a new understanding of the low-lying $N$ and $\Lambda$ resonances as illustrated in Fig.~\ref{fg:newp}.
With regard to the simple constituent quark model, we now understand that it predicts three levels
of hadron mass with approximately equal spacings.  It describes the ground state, an odd-parity
state the order of 500 MeV above the ground state and an even parity excitation approximately 1 GeV
above the ground state.

Within the context of a constituent quark model, a mass of 2 GeV for the first radial excitation is
natural.  With the first negative-parity excitation of the nucleon, the $N^\star(1535)$,
approximately 500 MeV above the nucleon, expectations -- based upon a phenomenologically successful
harmonic oscillator model -- suggest that the first positive-parity excited state should occur
another 500 MeV above, at around 2 GeV.

We now understand that the structure of the $\Lambda^*(1405)$ and $N^*(1440)$ is more complicated
and beyond the scope of a simple model based on three constituent quarks.
Through the analyses of Lattice QCD results with HEFT, the nature of the $\Lambda^*(1405)$ and
$N^*(1440)$ appear to be predominantly meson-baryon states.
In other words, these two states are beyond the quark model and tuning quark model parameters to
encompass these states spoils the insight to be gained from these models.
Indeed, the states predicted in the quark model as discussed above, are observed in the spectra of
contemporary lattice QCD results.

\section{Conclusion}
\label{sec:con}

We have examined experimental pion-nucleon scattering data in Hamiltonian effective field theory,
an extension of chiral perturbation theory that incorporates the L\"uscher relation connecting
the scattering phase shifts to the energy levels observed in finite volume lattice QCD
calculations.
By considering the Hamiltonian matrix in the finite-volume, one not only learns the energy
eigenvalues corresponding to the spectrum of Lattice QCD simulations, but one also learns the
eigenvectors of these eigenstates describing the composition of the states.
This information has been key to advancing our understanding of the structure of low-lying baryon
resonances.

We developed two scenarios for describing the Roper resonance in the pion-nucleon scattering
process.  Scenario I describes the Roper resonance as a baryon-meson scattering state, while
Scenario II describes the Roper as a traditional three-quark state.  Both scenarios can fit
the experimental data very well.  However, only Scenario I is consistent with lattice QCD results. 

A particular focus of this investigation has been to ascertain the possible role of a basis state
in the Hamiltonian formulation that can be associated with a quark-model-like radial excitation of
the nucleon.
In conclusion, the quark-model-like basis state associated with the Roper resonance lies at
approximately 2 GeV.  This large mass is required in order to provide a finite-volume spectrum
consistent with lattice QCD.  
A lower bare-basis-state mass leads to the prediction of quark-model
like low-lying states dominated by a bare-state component.  The absence of such states in the
lattice QCD spectrum rules out this scenario.

The description of both the experimental scattering data in Fig.~\ref{fg:phase} and the lattice
results presented in Fig.~\ref{fg:specI} for this scenario is excellent.  All localized lattice QCD
states are associated with a HEFT spectral line whose composition includes a large quark-model-like
basis state component.  Similarly, HEFT accurately predicts the positions of the scattering states
observed in lattice QCD by Lang {\it et al.}\ \cite{Lang:2016hnn} as well as their composition.

In the preferred Scenario I with a 2 GeV bare-state mass, quark-model-like states sit high in the
spectrum.  In the finite volume of the $L=3$ fm lattice, the bare state is dressed to produce states
commencing at $\simeq 1.9$ GeV for the lightest pion mass of 156 MeV.  Indeed, the CSSM studied the
three-quark wave function of this state and discovered it resembles the first radial excitation of
the quark model \cite{Roberts:2013ipa,Roberts:2013oea}.  In the infinite volume, this state is
associated with a pole at approximately 2 GeV.

Thus, it is now clear that there is an unconventional role for a quark-model-like description of
the Roper in describing experimental data and the lattice QCD results.  The analysis shows that one
can admit such a state provided it sits high in the spectrum at approximately 2 GeV.  Remarkably,
this is where simple quark models naturally place the radial excitation of the nucleon.

Even though there are no localized states seen on the lattice between this energy scale of $\simeq
1.9$ GeV and the ground state nucleon mass \cite{Kiratidis:2016hda}, Scenario I generates an
infinite-volume pole in the Roper resonance region of the spectrum.  This pole arises from strong
rescattering in the coupled meson-baryon channels, which dominate the underlying
structure of the Roper-resonance.

These conclusions reveal that the spectrum of quark-model-like states is relatively simple, once
one excludes the more exotic $\Lambda^*(1405)$ and $N^*(1440)$ from the traditional scope of the
model.
The $\Lambda^*(1405)$ and $N^*(1440)$ contain nontrivial meson-baryon interactions with the
$\Lambda^*(1405)$ dominated by a molecular $\bar K N$ component and the $N^*(1440)$ arising out of
strong meson-baryon rescattering.

Finally, the relationship between the radial excitation of the nucleon and the Roper resonance is
now understood.  The quark-model-like basis state at approximately 2 GeV makes a small but
nontrivial contribution to the finite volume state observed in the regime of the Roper resonance by
Lang {\it et al.} at $\simeq 1.5$ GeV.  While this state is excited with the momentum-projected
$\sigma N$ interpolating field, it has a small 2 GeV quark-model-like basis-state component in
HEFT, as indicated by the green curve in Fig.~\ref{fg:specI} through the second excitation reported
by Lang {\it et al.} \cite{Lang:2016hnn}.  This component may become the dominant component in
large $Q^2$ transition form factors \cite{Wilson:2011aa} where long-distance meson-cloud effects
are highly suppressed.  Still, the predominant structure of the Roper resonance has its origin in the
strong rescattering of $\pi N$, $\pi \Delta$ and $\sigma N$ channels.

With this new insight, the mystery of the low-lying Roper resonance may be nearing resolution.
Evidence indicates the observed nucleon resonance at 1440 MeV is best described as the result of
strong rescattering between coupled meson-baryon channels.

In working towards a definitive analysis there is ample scope for new data to further resolve the
nature of this state.  Further development of three-body channel
contributions~\cite{Hansen:2014eka, Hansen:2015zga} in effective field theory is desired.
Similarly, a more comprehensive lattice QCD analysis of the {\em complete} nucleon spectrum in
several lattice volumes would serve well to further expose the role of the coupled channels giving
rise to the Roper resonance.

\begin{acknowledgments}
{\bf Acknowledgements:} This research was supported by the Fundamental Research Funds 
of Lanzhou University under Grants 223000-862637 and by the Australian Research Council
through the ARC Centre of Excellence for Particle Physics at the
Terascale (CE110001104) and through Grants No.\ DP151103101 (A.W.T.),
DP140103067 and DP150103164 (D.B.L.).
\end{acknowledgments}


\bibliographystyle{apsrev4-1}
\bibliography{refs}

\begin{thebibliography}{57}%
\makeatletter
\providecommand \@ifxundefined [1]{%
 \@ifx{#1\undefined}
}%
\providecommand \@ifnum [1]{%
 \ifnum #1\expandafter \@firstoftwo
 \else \expandafter \@secondoftwo
 \fi
}%
\providecommand \@ifx [1]{%
 \ifx #1\expandafter \@firstoftwo
 \else \expandafter \@secondoftwo
 \fi
}%
\providecommand \natexlab [1]{#1}%
\providecommand \enquote  [1]{``#1''}%
\providecommand \bibnamefont  [1]{#1}%
\providecommand \bibfnamefont [1]{#1}%
\providecommand \citenamefont [1]{#1}%
\providecommand \href@noop [0]{\@secondoftwo}%
\providecommand \href [0]{\begingroup \@sanitize@url \@href}%
\providecommand \@href[1]{\@@startlink{#1}\@@href}%
\providecommand \@@href[1]{\endgroup#1\@@endlink}%
\providecommand \@sanitize@url [0]{\catcode `\\12\catcode `\$12\catcode
  `\&12\catcode `\#12\catcode `\^12\catcode `\_12\catcode `\%12\relax}%
\providecommand \@@startlink[1]{}%
\providecommand \@@endlink[0]{}%
\providecommand \url  [0]{\begingroup\@sanitize@url \@url }%
\providecommand \@url [1]{\endgroup\@href {#1}{\urlprefix }}%
\providecommand \urlprefix  [0]{URL }%
\providecommand \Eprint [0]{\href }%
\providecommand \doibase [0]{http://dx.doi.org/}%
\providecommand \selectlanguage [0]{\@gobble}%
\providecommand \bibinfo  [0]{\@secondoftwo}%
\providecommand \bibfield  [0]{\@secondoftwo}%
\providecommand \translation [1]{[#1]}%
\providecommand \BibitemOpen [0]{}%
\providecommand \bibitemStop [0]{}%
\providecommand \bibitemNoStop [0]{.\EOS\space}%
\providecommand \EOS [0]{\spacefactor3000\relax}%
\providecommand \BibitemShut  [1]{\csname bibitem#1\endcsname}%
\let\auto@bib@innerbib\@empty
\bibitem [{\citenamefont {Roper}(1964)}]{Roper:1964zza}%
  \BibitemOpen
  \bibfield  {author} {\bibinfo {author} {\bibfnamefont {L.~D.}\ \bibnamefont
  {Roper}},\ }\href {\doibase 10.1103/PhysRevLett.12.340} {\bibfield  {journal}
  {\bibinfo  {journal} {Phys. Rev. Lett.}\ }\textbf {\bibinfo {volume} {12}},\
  \bibinfo {pages} {340} (\bibinfo {year} {1964})}\BibitemShut {NoStop}%
\bibitem [{\citenamefont {Isgur}\ and\ \citenamefont
  {Karl}(1979)}]{Isgur:1978wd}%
  \BibitemOpen
  \bibfield  {author} {\bibinfo {author} {\bibfnamefont {N.}~\bibnamefont
  {Isgur}}\ and\ \bibinfo {author} {\bibfnamefont {G.}~\bibnamefont {Karl}},\
  }\href {\doibase 10.1103/PhysRevD.23.817.2, 10.1103/PhysRevD.19.2653}
  {\bibfield  {journal} {\bibinfo  {journal} {Phys. Rev.}\ }\textbf {\bibinfo
  {volume} {D19}},\ \bibinfo {pages} {2653} (\bibinfo {year} {1979})},\
  \bibinfo {note} {[Erratum: Phys. Rev.D23,817(1981)]}\BibitemShut {NoStop}%
\bibitem [{\citenamefont {Aznauryan}\ \emph {et~al.}(2008)\citenamefont
  {Aznauryan} \emph {et~al.}}]{Aznauryan:2008pe}%
  \BibitemOpen
  \bibfield  {author} {\bibinfo {author} {\bibfnamefont {I.~G.}\ \bibnamefont
  {Aznauryan}} \emph {et~al.} (\bibinfo {collaboration} {CLAS}),\ }\href
  {\doibase 10.1103/PhysRevC.78.045209} {\bibfield  {journal} {\bibinfo
  {journal} {Phys. Rev.}\ }\textbf {\bibinfo {volume} {C78}},\ \bibinfo {pages}
  {045209} (\bibinfo {year} {2008})},\ \Eprint {http://arxiv.org/abs/0804.0447}
  {arXiv:0804.0447 [nucl-ex]} \BibitemShut {NoStop}%
\bibitem [{\citenamefont {Joo}\ \emph {et~al.}(2005)\citenamefont {Joo} \emph
  {et~al.}}]{Joo:2005gs}%
  \BibitemOpen
  \bibfield  {author} {\bibinfo {author} {\bibfnamefont {K.}~\bibnamefont
  {Joo}} \emph {et~al.} (\bibinfo {collaboration} {CLAS}),\ }\href {\doibase
  10.1103/PhysRevC.72.058202} {\bibfield  {journal} {\bibinfo  {journal} {Phys.
  Rev.}\ }\textbf {\bibinfo {volume} {C72}},\ \bibinfo {pages} {058202}
  (\bibinfo {year} {2005})},\ \Eprint {http://arxiv.org/abs/nucl-ex/0504027}
  {arXiv:nucl-ex/0504027 [nucl-ex]} \BibitemShut {NoStop}%
\bibitem [{\citenamefont {Weber}(1990)}]{Weber:1989fv}%
  \BibitemOpen
  \bibfield  {author} {\bibinfo {author} {\bibfnamefont {H.~J.}\ \bibnamefont
  {Weber}},\ }\href {\doibase 10.1103/PhysRevC.41.2783} {\bibfield  {journal}
  {\bibinfo  {journal} {Phys. Rev.}\ }\textbf {\bibinfo {volume} {C41}},\
  \bibinfo {pages} {2783} (\bibinfo {year} {1990})}\BibitemShut {NoStop}%
\bibitem [{\citenamefont {Julia-Diaz}\ and\ \citenamefont
  {Riska}(2006)}]{JuliaDiaz:2006av}%
  \BibitemOpen
  \bibfield  {author} {\bibinfo {author} {\bibfnamefont {B.}~\bibnamefont
  {Julia-Diaz}}\ and\ \bibinfo {author} {\bibfnamefont {D.~O.}\ \bibnamefont
  {Riska}},\ }\href {\doibase 10.1016/j.nuclphysa.2006.09.016} {\bibfield
  {journal} {\bibinfo  {journal} {Nucl. Phys.}\ }\textbf {\bibinfo {volume}
  {A780}},\ \bibinfo {pages} {175} (\bibinfo {year} {2006})},\ \Eprint
  {http://arxiv.org/abs/nucl-th/0609064} {arXiv:nucl-th/0609064 [nucl-th]}
  \BibitemShut {NoStop}%
\bibitem [{\citenamefont {Barquilla-Cano}\ \emph {et~al.}(2007)\citenamefont
  {Barquilla-Cano}, \citenamefont {Buchmann},\ and\ \citenamefont
  {Hernandez}}]{BarquillaCano:2007yk}%
  \BibitemOpen
  \bibfield  {author} {\bibinfo {author} {\bibfnamefont {D.}~\bibnamefont
  {Barquilla-Cano}}, \bibinfo {author} {\bibfnamefont {A.~J.}\ \bibnamefont
  {Buchmann}}, \ and\ \bibinfo {author} {\bibfnamefont {E.}~\bibnamefont
  {Hernandez}},\ }\href {\doibase 10.1103/PhysRevC.77.019903,
  10.1103/PhysRevC.75.065203} {\bibfield  {journal} {\bibinfo  {journal} {Phys.
  Rev.}\ }\textbf {\bibinfo {volume} {C75}},\ \bibinfo {pages} {065203}
  (\bibinfo {year} {2007})},\ \bibinfo {note} {[Erratum: Phys.
  Rev.C77,019903(2008)]},\ \Eprint {http://arxiv.org/abs/0705.3297}
  {arXiv:0705.3297 [nucl-th]} \BibitemShut {NoStop}%
\bibitem [{\citenamefont {Golli}\ and\ \citenamefont
  {Sirca}(2008)}]{Golli:2007sa}%
  \BibitemOpen
  \bibfield  {author} {\bibinfo {author} {\bibfnamefont {B.}~\bibnamefont
  {Golli}}\ and\ \bibinfo {author} {\bibfnamefont {S.}~\bibnamefont {Sirca}},\
  }\href {\doibase 10.1140/epja/i2008-10677-3} {\bibfield  {journal} {\bibinfo
  {journal} {Eur. Phys. J.}\ }\textbf {\bibinfo {volume} {A38}},\ \bibinfo
  {pages} {271} (\bibinfo {year} {2008})},\ \Eprint
  {http://arxiv.org/abs/0708.3759} {arXiv:0708.3759 [hep-ph]} \BibitemShut
  {NoStop}%
\bibitem [{\citenamefont {Golli}\ \emph {et~al.}(2009)\citenamefont {Golli},
  \citenamefont {Sirca},\ and\ \citenamefont {Fiolhais}}]{Golli:2009uk}%
  \BibitemOpen
  \bibfield  {author} {\bibinfo {author} {\bibfnamefont {B.}~\bibnamefont
  {Golli}}, \bibinfo {author} {\bibfnamefont {S.}~\bibnamefont {Sirca}}, \ and\
  \bibinfo {author} {\bibfnamefont {M.}~\bibnamefont {Fiolhais}},\ }\href
  {\doibase 10.1140/epja/i2009-10878-2} {\bibfield  {journal} {\bibinfo
  {journal} {Eur. Phys. J.}\ }\textbf {\bibinfo {volume} {A42}},\ \bibinfo
  {pages} {185} (\bibinfo {year} {2009})},\ \Eprint
  {http://arxiv.org/abs/0906.2066} {arXiv:0906.2066 [nucl-th]} \BibitemShut
  {NoStop}%
\bibitem [{\citenamefont {Meissner}\ and\ \citenamefont
  {Durso}(1984)}]{Meissner:1984un}%
  \BibitemOpen
  \bibfield  {author} {\bibinfo {author} {\bibfnamefont {U.~G.}\ \bibnamefont
  {Meissner}}\ and\ \bibinfo {author} {\bibfnamefont {J.~W.}\ \bibnamefont
  {Durso}},\ }\href {\doibase 10.1016/0375-9474(84)90100-3} {\bibfield
  {journal} {\bibinfo  {journal} {Nucl. Phys.}\ }\textbf {\bibinfo {volume}
  {A430}},\ \bibinfo {pages} {670} (\bibinfo {year} {1984})}\BibitemShut
  {NoStop}%
\bibitem [{\citenamefont {Hajduk}\ and\ \citenamefont
  {Schwesinger}(1984)}]{Hajduk:1984ry}%
  \BibitemOpen
  \bibfield  {author} {\bibinfo {author} {\bibfnamefont {C.}~\bibnamefont
  {Hajduk}}\ and\ \bibinfo {author} {\bibfnamefont {B.}~\bibnamefont
  {Schwesinger}},\ }\href {\doibase 10.1016/0370-2693(84)90914-6} {\bibfield
  {journal} {\bibinfo  {journal} {Phys. Lett.}\ }\textbf {\bibinfo {volume}
  {B140}},\ \bibinfo {pages} {172} (\bibinfo {year} {1984})}\BibitemShut
  {NoStop}%
\bibitem [{\citenamefont {Krehl}\ \emph {et~al.}(2000)\citenamefont {Krehl},
  \citenamefont {Hanhart}, \citenamefont {Krewald},\ and\ \citenamefont
  {Speth}}]{Krehl:1999km}%
  \BibitemOpen
  \bibfield  {author} {\bibinfo {author} {\bibfnamefont {O.}~\bibnamefont
  {Krehl}}, \bibinfo {author} {\bibfnamefont {C.}~\bibnamefont {Hanhart}},
  \bibinfo {author} {\bibfnamefont {S.}~\bibnamefont {Krewald}}, \ and\
  \bibinfo {author} {\bibfnamefont {J.}~\bibnamefont {Speth}},\ }\href
  {\doibase 10.1103/PhysRevC.62.025207} {\bibfield  {journal} {\bibinfo
  {journal} {Phys. Rev.}\ }\textbf {\bibinfo {volume} {C62}},\ \bibinfo {pages}
  {025207} (\bibinfo {year} {2000})},\ \Eprint
  {http://arxiv.org/abs/nucl-th/9911080} {arXiv:nucl-th/9911080 [nucl-th]}
  \BibitemShut {NoStop}%
\bibitem [{\citenamefont {Schutz}\ \emph {et~al.}(1998)\citenamefont {Schutz},
  \citenamefont {Haidenbauer}, \citenamefont {Speth},\ and\ \citenamefont
  {Durso}}]{Schutz:1998jx}%
  \BibitemOpen
  \bibfield  {author} {\bibinfo {author} {\bibfnamefont {C.}~\bibnamefont
  {Schutz}}, \bibinfo {author} {\bibfnamefont {J.}~\bibnamefont {Haidenbauer}},
  \bibinfo {author} {\bibfnamefont {J.}~\bibnamefont {Speth}}, \ and\ \bibinfo
  {author} {\bibfnamefont {J.~W.}\ \bibnamefont {Durso}},\ }\href {\doibase
  10.1103/PhysRevC.57.1464} {\bibfield  {journal} {\bibinfo  {journal} {Phys.
  Rev.}\ }\textbf {\bibinfo {volume} {C57}},\ \bibinfo {pages} {1464} (\bibinfo
  {year} {1998})}\BibitemShut {NoStop}%
\bibitem [{\citenamefont {Matsuyama}\ \emph {et~al.}(2007)\citenamefont
  {Matsuyama}, \citenamefont {Sato},\ and\ \citenamefont
  {Lee}}]{Matsuyama:2006rp}%
  \BibitemOpen
  \bibfield  {author} {\bibinfo {author} {\bibfnamefont {A.}~\bibnamefont
  {Matsuyama}}, \bibinfo {author} {\bibfnamefont {T.}~\bibnamefont {Sato}}, \
  and\ \bibinfo {author} {\bibfnamefont {T.~S.~H.}\ \bibnamefont {Lee}},\
  }\href {\doibase 10.1016/j.physrep.2006.12.003} {\bibfield  {journal}
  {\bibinfo  {journal} {Phys. Rept.}\ }\textbf {\bibinfo {volume} {439}},\
  \bibinfo {pages} {193} (\bibinfo {year} {2007})},\ \Eprint
  {http://arxiv.org/abs/nucl-th/0608051} {arXiv:nucl-th/0608051 [nucl-th]}
  \BibitemShut {NoStop}%
\bibitem [{\citenamefont {Kamano}\ \emph {et~al.}(2010)\citenamefont {Kamano},
  \citenamefont {Nakamura}, \citenamefont {Lee},\ and\ \citenamefont
  {Sato}}]{Kamano:2010ud}%
  \BibitemOpen
  \bibfield  {author} {\bibinfo {author} {\bibfnamefont {H.}~\bibnamefont
  {Kamano}}, \bibinfo {author} {\bibfnamefont {S.~X.}\ \bibnamefont
  {Nakamura}}, \bibinfo {author} {\bibfnamefont {T.~S.~H.}\ \bibnamefont
  {Lee}}, \ and\ \bibinfo {author} {\bibfnamefont {T.}~\bibnamefont {Sato}},\
  }\href {\doibase 10.1103/PhysRevC.81.065207} {\bibfield  {journal} {\bibinfo
  {journal} {Phys. Rev.}\ }\textbf {\bibinfo {volume} {C81}},\ \bibinfo {pages}
  {065207} (\bibinfo {year} {2010})},\ \Eprint {http://arxiv.org/abs/1001.5083}
  {arXiv:1001.5083 [nucl-th]} \BibitemShut {NoStop}%
\bibitem [{\citenamefont {Kamano}\ \emph {et~al.}(2013)\citenamefont {Kamano},
  \citenamefont {Nakamura}, \citenamefont {Lee},\ and\ \citenamefont
  {Sato}}]{Kamano:2013iva}%
  \BibitemOpen
  \bibfield  {author} {\bibinfo {author} {\bibfnamefont {H.}~\bibnamefont
  {Kamano}}, \bibinfo {author} {\bibfnamefont {S.~X.}\ \bibnamefont
  {Nakamura}}, \bibinfo {author} {\bibfnamefont {T.~S.~H.}\ \bibnamefont
  {Lee}}, \ and\ \bibinfo {author} {\bibfnamefont {T.}~\bibnamefont {Sato}},\
  }\href {\doibase 10.1103/PhysRevC.88.035209} {\bibfield  {journal} {\bibinfo
  {journal} {Phys. Rev.}\ }\textbf {\bibinfo {volume} {C88}},\ \bibinfo {pages}
  {035209} (\bibinfo {year} {2013})},\ \Eprint {http://arxiv.org/abs/1305.4351}
  {arXiv:1305.4351 [nucl-th]} \BibitemShut {NoStop}%
\bibitem [{\citenamefont {Suzuki}\ \emph {et~al.}(2010)\citenamefont {Suzuki},
  \citenamefont {Julia-Diaz}, \citenamefont {Kamano}, \citenamefont {Lee},
  \citenamefont {Matsuyama},\ and\ \citenamefont {Sato}}]{Suzuki:2009nj}%
  \BibitemOpen
  \bibfield  {author} {\bibinfo {author} {\bibfnamefont {N.}~\bibnamefont
  {Suzuki}}, \bibinfo {author} {\bibfnamefont {B.}~\bibnamefont {Julia-Diaz}},
  \bibinfo {author} {\bibfnamefont {H.}~\bibnamefont {Kamano}}, \bibinfo
  {author} {\bibfnamefont {T.~S.~H.}\ \bibnamefont {Lee}}, \bibinfo {author}
  {\bibfnamefont {A.}~\bibnamefont {Matsuyama}}, \ and\ \bibinfo {author}
  {\bibfnamefont {T.}~\bibnamefont {Sato}},\ }\href {\doibase
  10.1103/PhysRevLett.104.042302} {\bibfield  {journal} {\bibinfo  {journal}
  {Phys. Rev. Lett.}\ }\textbf {\bibinfo {volume} {104}},\ \bibinfo {pages}
  {042302} (\bibinfo {year} {2010})},\ \Eprint {http://arxiv.org/abs/0909.1356}
  {arXiv:0909.1356 [nucl-th]} \BibitemShut {NoStop}%
\bibitem [{\citenamefont {Hernandez}\ \emph {et~al.}(2002)\citenamefont
  {Hernandez}, \citenamefont {Oset},\ and\ \citenamefont
  {Vicente~Vacas}}]{Hernandez:2002xk}%
  \BibitemOpen
  \bibfield  {author} {\bibinfo {author} {\bibfnamefont {E.}~\bibnamefont
  {Hernandez}}, \bibinfo {author} {\bibfnamefont {E.}~\bibnamefont {Oset}}, \
  and\ \bibinfo {author} {\bibfnamefont {M.~J.}\ \bibnamefont
  {Vicente~Vacas}},\ }\href {\doibase 10.1103/PhysRevC.66.065201} {\bibfield
  {journal} {\bibinfo  {journal} {Phys. Rev.}\ }\textbf {\bibinfo {volume}
  {C66}},\ \bibinfo {pages} {065201} (\bibinfo {year} {2002})},\ \Eprint
  {http://arxiv.org/abs/nucl-th/0209009} {arXiv:nucl-th/0209009 [nucl-th]}
  \BibitemShut {NoStop}%
\bibitem [{\citenamefont {Barnes}\ and\ \citenamefont
  {Close}(1983)}]{Barnes:1982fj}%
  \BibitemOpen
  \bibfield  {author} {\bibinfo {author} {\bibfnamefont {T.}~\bibnamefont
  {Barnes}}\ and\ \bibinfo {author} {\bibfnamefont {F.~E.}\ \bibnamefont
  {Close}},\ }\href {\doibase 10.1016/0370-2693(83)90965-6} {\bibfield
  {journal} {\bibinfo  {journal} {Phys. Lett.}\ }\textbf {\bibinfo {volume}
  {B123}},\ \bibinfo {pages} {89} (\bibinfo {year} {1983})}\BibitemShut
  {NoStop}%
\bibitem [{\citenamefont {Golowich}\ \emph {et~al.}(1983)\citenamefont
  {Golowich}, \citenamefont {Haqq},\ and\ \citenamefont
  {Karl}}]{Golowich:1982kx}%
  \BibitemOpen
  \bibfield  {author} {\bibinfo {author} {\bibfnamefont {E.}~\bibnamefont
  {Golowich}}, \bibinfo {author} {\bibfnamefont {E.}~\bibnamefont {Haqq}}, \
  and\ \bibinfo {author} {\bibfnamefont {G.}~\bibnamefont {Karl}},\ }\href
  {\doibase 10.1103/PhysRevD.28.160, 10.1103/PhysRevD.33.859} {\bibfield
  {journal} {\bibinfo  {journal} {Phys. Rev.}\ }\textbf {\bibinfo {volume}
  {D28}},\ \bibinfo {pages} {160} (\bibinfo {year} {1983})},\ \bibinfo {note}
  {[Erratum: Phys. Rev.D33,859(1986)]}\BibitemShut {NoStop}%
\bibitem [{\citenamefont {Kisslinger}\ and\ \citenamefont
  {Li}(1995)}]{Kisslinger:1995yw}%
  \BibitemOpen
  \bibfield  {author} {\bibinfo {author} {\bibfnamefont {L.~S.}\ \bibnamefont
  {Kisslinger}}\ and\ \bibinfo {author} {\bibfnamefont {Z.~P.}\ \bibnamefont
  {Li}},\ }\href {\doibase 10.1103/PhysRevD.51.R5986} {\bibfield  {journal}
  {\bibinfo  {journal} {Phys. Rev.}\ }\textbf {\bibinfo {volume} {D51}},\
  \bibinfo {pages} {R5986} (\bibinfo {year} {1995})}\BibitemShut {NoStop}%
\bibitem [{\citenamefont {Patrignani}\ \emph {et~al.}(2016)\citenamefont
  {Patrignani} \emph {et~al.}}]{Olive:2016xmw}%
  \BibitemOpen
  \bibfield  {author} {\bibinfo {author} {\bibfnamefont {C.}~\bibnamefont
  {Patrignani}} \emph {et~al.} (\bibinfo {collaboration} {Particle Data
  Group}),\ }\href {\doibase 10.1088/1674-1137/40/10/100001} {\bibfield
  {journal} {\bibinfo  {journal} {Chin. Phys.}\ }\textbf {\bibinfo {volume}
  {C40}},\ \bibinfo {pages} {100001} (\bibinfo {year} {2016})}\BibitemShut
  {NoStop}%
\bibitem [{\citenamefont {Hall}\ \emph {et~al.}(2015)\citenamefont {Hall},
  \citenamefont {Kamleh}, \citenamefont {Leinweber}, \citenamefont {Menadue},
  \citenamefont {Owen}, \citenamefont {Thomas},\ and\ \citenamefont
  {Young}}]{Hall:2014uca}%
  \BibitemOpen
  \bibfield  {author} {\bibinfo {author} {\bibfnamefont {J.~M.~M.}\
  \bibnamefont {Hall}}, \bibinfo {author} {\bibfnamefont {W.}~\bibnamefont
  {Kamleh}}, \bibinfo {author} {\bibfnamefont {D.~B.}\ \bibnamefont
  {Leinweber}}, \bibinfo {author} {\bibfnamefont {B.~J.}\ \bibnamefont
  {Menadue}}, \bibinfo {author} {\bibfnamefont {B.~J.}\ \bibnamefont {Owen}},
  \bibinfo {author} {\bibfnamefont {A.~W.}\ \bibnamefont {Thomas}}, \ and\
  \bibinfo {author} {\bibfnamefont {R.~D.}\ \bibnamefont {Young}},\ }\href
  {\doibase 10.1103/PhysRevLett.114.132002} {\bibfield  {journal} {\bibinfo
  {journal} {Phys. Rev. Lett.}\ }\textbf {\bibinfo {volume} {114}},\ \bibinfo
  {pages} {132002} (\bibinfo {year} {2015})},\ \Eprint
  {http://arxiv.org/abs/1411.3402} {arXiv:1411.3402 [hep-lat]} \BibitemShut
  {NoStop}%
\bibitem [{\citenamefont {Hall}\ \emph {et~al.}(2017)\citenamefont {Hall},
  \citenamefont {Kamleh}, \citenamefont {Leinweber}, \citenamefont {Menadue},
  \citenamefont {Owen},\ and\ \citenamefont {Thomas}}]{Hall:2016kou}%
  \BibitemOpen
  \bibfield  {author} {\bibinfo {author} {\bibfnamefont {J.~M.~M.}\
  \bibnamefont {Hall}}, \bibinfo {author} {\bibfnamefont {W.}~\bibnamefont
  {Kamleh}}, \bibinfo {author} {\bibfnamefont {D.~B.}\ \bibnamefont
  {Leinweber}}, \bibinfo {author} {\bibfnamefont {B.~J.}\ \bibnamefont
  {Menadue}}, \bibinfo {author} {\bibfnamefont {B.~J.}\ \bibnamefont {Owen}}, \
  and\ \bibinfo {author} {\bibfnamefont {A.~W.}\ \bibnamefont {Thomas}},\
  }\href {\doibase 10.1103/PhysRevD.95.054510} {\bibfield  {journal} {\bibinfo
  {journal} {Phys. Rev.}\ }\textbf {\bibinfo {volume} {D95}},\ \bibinfo {pages}
  {054510} (\bibinfo {year} {2017})},\ \Eprint
  {http://arxiv.org/abs/1612.07477} {arXiv:1612.07477 [hep-lat]} \BibitemShut
  {NoStop}%
\bibitem [{\citenamefont {Liu}\ \emph {et~al.}(2016)\citenamefont {Liu},
  \citenamefont {Kamleh}, \citenamefont {Leinweber}, \citenamefont {Stokes},
  \citenamefont {Thomas},\ and\ \citenamefont {Wu}}]{Liu:2015ktc}%
  \BibitemOpen
  \bibfield  {author} {\bibinfo {author} {\bibfnamefont {Z.-W.}\ \bibnamefont
  {Liu}}, \bibinfo {author} {\bibfnamefont {W.}~\bibnamefont {Kamleh}},
  \bibinfo {author} {\bibfnamefont {D.~B.}\ \bibnamefont {Leinweber}}, \bibinfo
  {author} {\bibfnamefont {F.~M.}\ \bibnamefont {Stokes}}, \bibinfo {author}
  {\bibfnamefont {A.~W.}\ \bibnamefont {Thomas}}, \ and\ \bibinfo {author}
  {\bibfnamefont {J.-J.}\ \bibnamefont {Wu}},\ }\href {\doibase
  10.1103/PhysRevLett.116.082004} {\bibfield  {journal} {\bibinfo  {journal}
  {Phys. Rev. Lett.}\ }\textbf {\bibinfo {volume} {116}},\ \bibinfo {pages}
  {082004} (\bibinfo {year} {2016})},\ \Eprint
  {http://arxiv.org/abs/1512.00140} {arXiv:1512.00140 [hep-lat]} \BibitemShut
  {NoStop}%
\bibitem [{\citenamefont {Liu}\ \emph {et~al.}(2017{\natexlab{a}})\citenamefont
  {Liu}, \citenamefont {Hall}, \citenamefont {Leinweber}, \citenamefont
  {Thomas},\ and\ \citenamefont {Wu}}]{Liu:2016wxq}%
  \BibitemOpen
  \bibfield  {author} {\bibinfo {author} {\bibfnamefont {Z.-W.}\ \bibnamefont
  {Liu}}, \bibinfo {author} {\bibfnamefont {J.~M.~M.}\ \bibnamefont {Hall}},
  \bibinfo {author} {\bibfnamefont {D.~B.}\ \bibnamefont {Leinweber}}, \bibinfo
  {author} {\bibfnamefont {A.~W.}\ \bibnamefont {Thomas}}, \ and\ \bibinfo
  {author} {\bibfnamefont {J.-J.}\ \bibnamefont {Wu}},\ }\href {\doibase
  10.1103/PhysRevD.95.014506} {\bibfield  {journal} {\bibinfo  {journal} {Phys.
  Rev.}\ }\textbf {\bibinfo {volume} {D95}},\ \bibinfo {pages} {014506}
  (\bibinfo {year} {2017}{\natexlab{a}})},\ \Eprint
  {http://arxiv.org/abs/1607.05856} {arXiv:1607.05856 [nucl-th]} \BibitemShut
  {NoStop}%
\bibitem [{\citenamefont {Molina}\ and\ \citenamefont
  {Döring}(2016)}]{Molina:2015uqp}%
  \BibitemOpen
  \bibfield  {author} {\bibinfo {author} {\bibfnamefont {R.}~\bibnamefont
  {Molina}}\ and\ \bibinfo {author} {\bibfnamefont {M.}~\bibnamefont
  {Döring}},\ }\href {\doibase 10.1103/PhysRevD.94.056010,
  10.1103/PhysRevD.94.079901} {\bibfield  {journal} {\bibinfo  {journal} {Phys.
  Rev.}\ }\textbf {\bibinfo {volume} {D94}},\ \bibinfo {pages} {056010}
  (\bibinfo {year} {2016})},\ \bibinfo {note} {[Addendum: Phys.
  Rev.D94,no.7,079901(2016)]},\ \Eprint {http://arxiv.org/abs/1512.05831}
  {arXiv:1512.05831 [hep-lat]} \BibitemShut {NoStop}%
\bibitem [{\citenamefont {Luscher}(1986)}]{Luscher:1986pf}%
  \BibitemOpen
  \bibfield  {author} {\bibinfo {author} {\bibfnamefont {M.}~\bibnamefont
  {Luscher}},\ }\href {\doibase 10.1007/BF01211097} {\bibfield  {journal}
  {\bibinfo  {journal} {Commun. Math. Phys.}\ }\textbf {\bibinfo {volume}
  {105}},\ \bibinfo {pages} {153} (\bibinfo {year} {1986})}\BibitemShut
  {NoStop}%
\bibitem [{\citenamefont {Luscher}(1991{\natexlab{a}})}]{Luscher:1990ux}%
  \BibitemOpen
  \bibfield  {author} {\bibinfo {author} {\bibfnamefont {M.}~\bibnamefont
  {Luscher}},\ }\href {\doibase 10.1016/0550-3213(91)90366-6} {\bibfield
  {journal} {\bibinfo  {journal} {Nucl. Phys.}\ }\textbf {\bibinfo {volume}
  {B354}},\ \bibinfo {pages} {531} (\bibinfo {year}
  {1991}{\natexlab{a}})}\BibitemShut {NoStop}%
\bibitem [{\citenamefont {Luscher}(1991{\natexlab{b}})}]{Luscher:1991cf}%
  \BibitemOpen
  \bibfield  {author} {\bibinfo {author} {\bibfnamefont {M.}~\bibnamefont
  {Luscher}},\ }\href {\doibase 10.1016/0550-3213(91)90584-K} {\bibfield
  {journal} {\bibinfo  {journal} {Nucl. Phys.}\ }\textbf {\bibinfo {volume}
  {B364}},\ \bibinfo {pages} {237} (\bibinfo {year}
  {1991}{\natexlab{b}})}\BibitemShut {NoStop}%
\bibitem [{\citenamefont {Wu}\ \emph {et~al.}(2014)\citenamefont {Wu},
  \citenamefont {Lee}, \citenamefont {Thomas},\ and\ \citenamefont
  {Young}}]{Wu:2014vma}%
  \BibitemOpen
  \bibfield  {author} {\bibinfo {author} {\bibfnamefont {J.-J.}\ \bibnamefont
  {Wu}}, \bibinfo {author} {\bibfnamefont {T.~S.~H.}\ \bibnamefont {Lee}},
  \bibinfo {author} {\bibfnamefont {A.~W.}\ \bibnamefont {Thomas}}, \ and\
  \bibinfo {author} {\bibfnamefont {R.~D.}\ \bibnamefont {Young}},\ }\href
  {\doibase 10.1103/PhysRevC.90.055206} {\bibfield  {journal} {\bibinfo
  {journal} {Phys. Rev.}\ }\textbf {\bibinfo {volume} {C90}},\ \bibinfo {pages}
  {055206} (\bibinfo {year} {2014})},\ \Eprint {http://arxiv.org/abs/1402.4868}
  {arXiv:1402.4868 [hep-lat]} \BibitemShut {NoStop}%
\bibitem [{\citenamefont {Young}\ \emph {et~al.}(2003)\citenamefont {Young},
  \citenamefont {Leinweber},\ and\ \citenamefont {Thomas}}]{Young:2002ib}%
  \BibitemOpen
  \bibfield  {author} {\bibinfo {author} {\bibfnamefont {R.~D.}\ \bibnamefont
  {Young}}, \bibinfo {author} {\bibfnamefont {D.~B.}\ \bibnamefont
  {Leinweber}}, \ and\ \bibinfo {author} {\bibfnamefont {A.~W.}\ \bibnamefont
  {Thomas}},\ }\bibfield  {booktitle} {\emph {\bibinfo {booktitle} {{Quarks in
  hadrons and nuclei. Proceedings, International School of Nuclear Physics,
  24th Course, Erice, Italy, September 16-24, 2002}}},\ }\href {\doibase
  10.1016/S0146-6410(03)00034-6} {\bibfield  {journal} {\bibinfo  {journal}
  {Prog. Part. Nucl. Phys.}\ }\textbf {\bibinfo {volume} {50}},\ \bibinfo
  {pages} {399} (\bibinfo {year} {2003})},\ \bibinfo {note} {[,399(2002)]},\
  \Eprint {http://arxiv.org/abs/hep-lat/0212031} {arXiv:hep-lat/0212031
  [hep-lat]} \BibitemShut {NoStop}%
\bibitem [{\citenamefont {Wilson}\ \emph {et~al.}(2012)\citenamefont {Wilson},
  \citenamefont {Cloet}, \citenamefont {Chang},\ and\ \citenamefont
  {Roberts}}]{Wilson:2011aa}%
  \BibitemOpen
  \bibfield  {author} {\bibinfo {author} {\bibfnamefont {D.~J.}\ \bibnamefont
  {Wilson}}, \bibinfo {author} {\bibfnamefont {I.~C.}\ \bibnamefont {Cloet}},
  \bibinfo {author} {\bibfnamefont {L.}~\bibnamefont {Chang}}, \ and\ \bibinfo
  {author} {\bibfnamefont {C.~D.}\ \bibnamefont {Roberts}},\ }\href {\doibase
  10.1103/PhysRevC.85.025205} {\bibfield  {journal} {\bibinfo  {journal} {Phys.
  Rev.}\ }\textbf {\bibinfo {volume} {C85}},\ \bibinfo {pages} {025205}
  (\bibinfo {year} {2012})},\ \Eprint {http://arxiv.org/abs/1112.2212}
  {arXiv:1112.2212 [nucl-th]} \BibitemShut {NoStop}%
\bibitem [{\citenamefont {Liu}\ \emph {et~al.}(2017{\natexlab{b}})\citenamefont
  {Liu}, \citenamefont {Kamleh}, \citenamefont {Leinweber}, \citenamefont
  {Stokes}, \citenamefont {Thomas},\ and\ \citenamefont {Wu}}]{Liu:2016uzk}%
  \BibitemOpen
  \bibfield  {author} {\bibinfo {author} {\bibfnamefont {Z.-W.}\ \bibnamefont
  {Liu}}, \bibinfo {author} {\bibfnamefont {W.}~\bibnamefont {Kamleh}},
  \bibinfo {author} {\bibfnamefont {D.~B.}\ \bibnamefont {Leinweber}}, \bibinfo
  {author} {\bibfnamefont {F.~M.}\ \bibnamefont {Stokes}}, \bibinfo {author}
  {\bibfnamefont {A.~W.}\ \bibnamefont {Thomas}}, \ and\ \bibinfo {author}
  {\bibfnamefont {J.-J.}\ \bibnamefont {Wu}},\ }\href {\doibase
  10.1103/PhysRevD.95.034034} {\bibfield  {journal} {\bibinfo  {journal} {Phys.
  Rev.}\ }\textbf {\bibinfo {volume} {D95}},\ \bibinfo {pages} {034034}
  (\bibinfo {year} {2017}{\natexlab{b}})},\ \Eprint
  {http://arxiv.org/abs/1607.04536} {arXiv:1607.04536 [nucl-th]} \BibitemShut
  {NoStop}%
\bibitem [{\citenamefont {Hall}\ \emph {et~al.}(2013)\citenamefont {Hall},
  \citenamefont {Hsu}, \citenamefont {Leinweber}, \citenamefont {Thomas},\ and\
  \citenamefont {Young}}]{Hall:2013qba}%
  \BibitemOpen
  \bibfield  {author} {\bibinfo {author} {\bibfnamefont {J.~M.~M.}\
  \bibnamefont {Hall}}, \bibinfo {author} {\bibfnamefont {A.~C.~P.}\
  \bibnamefont {Hsu}}, \bibinfo {author} {\bibfnamefont {D.~B.}\ \bibnamefont
  {Leinweber}}, \bibinfo {author} {\bibfnamefont {A.~W.}\ \bibnamefont
  {Thomas}}, \ and\ \bibinfo {author} {\bibfnamefont {R.~D.}\ \bibnamefont
  {Young}},\ }\href {\doibase 10.1103/PhysRevD.87.094510} {\bibfield  {journal}
  {\bibinfo  {journal} {Phys. Rev.}\ }\textbf {\bibinfo {volume} {D87}},\
  \bibinfo {pages} {094510} (\bibinfo {year} {2013})},\ \Eprint
  {http://arxiv.org/abs/1303.4157} {arXiv:1303.4157 [hep-lat]} \BibitemShut
  {NoStop}%
\bibitem [{\citenamefont {Wu}\ \emph {et~al.}(2017)\citenamefont {Wu},
  \citenamefont {Kamano}, \citenamefont {Lee}, \citenamefont {Leinweber},\ and\
  \citenamefont {Thomas}}]{Wu:2016ixr}%
  \BibitemOpen
  \bibfield  {author} {\bibinfo {author} {\bibfnamefont {J.-J.}\ \bibnamefont
  {Wu}}, \bibinfo {author} {\bibfnamefont {H.}~\bibnamefont {Kamano}}, \bibinfo
  {author} {\bibfnamefont {T.~S.~H.}\ \bibnamefont {Lee}}, \bibinfo {author}
  {\bibfnamefont {D.~B.}\ \bibnamefont {Leinweber}}, \ and\ \bibinfo {author}
  {\bibfnamefont {A.~W.}\ \bibnamefont {Thomas}},\ }\href {\doibase
  10.1103/PhysRevD.95.114507} {\bibfield  {journal} {\bibinfo  {journal} {Phys.
  Rev.}\ }\textbf {\bibinfo {volume} {D95}},\ \bibinfo {pages} {114507}
  (\bibinfo {year} {2017})},\ \Eprint {http://arxiv.org/abs/1611.05970}
  {arXiv:1611.05970 [hep-lat]} \BibitemShut {NoStop}%
\bibitem [{\citenamefont {Cloet}\ \emph {et~al.}(2002)\citenamefont {Cloet},
  \citenamefont {Leinweber},\ and\ \citenamefont {Thomas}}]{Cloet:2002eg}%
  \BibitemOpen
  \bibfield  {author} {\bibinfo {author} {\bibfnamefont {I.~C.}\ \bibnamefont
  {Cloet}}, \bibinfo {author} {\bibfnamefont {D.~B.}\ \bibnamefont
  {Leinweber}}, \ and\ \bibinfo {author} {\bibfnamefont {A.~W.}\ \bibnamefont
  {Thomas}},\ }\href {\doibase 10.1103/PhysRevC.65.062201} {\bibfield
  {journal} {\bibinfo  {journal} {Phys. Rev.}\ }\textbf {\bibinfo {volume}
  {C65}},\ \bibinfo {pages} {062201} (\bibinfo {year} {2002})},\ \Eprint
  {http://arxiv.org/abs/hep-ph/0203023} {arXiv:hep-ph/0203023 [hep-ph]}
  \BibitemShut {NoStop}%
\bibitem [{\citenamefont {Julia-Diaz}\ \emph {et~al.}(2007)\citenamefont
  {Julia-Diaz}, \citenamefont {Lee}, \citenamefont {Matsuyama},\ and\
  \citenamefont {Sato}}]{JuliaDiaz:2007kz}%
  \BibitemOpen
  \bibfield  {author} {\bibinfo {author} {\bibfnamefont {B.}~\bibnamefont
  {Julia-Diaz}}, \bibinfo {author} {\bibfnamefont {T.~S.~H.}\ \bibnamefont
  {Lee}}, \bibinfo {author} {\bibfnamefont {A.}~\bibnamefont {Matsuyama}}, \
  and\ \bibinfo {author} {\bibfnamefont {T.}~\bibnamefont {Sato}},\ }\href
  {\doibase 10.1103/PhysRevC.76.065201} {\bibfield  {journal} {\bibinfo
  {journal} {Phys. Rev.}\ }\textbf {\bibinfo {volume} {C76}},\ \bibinfo {pages}
  {065201} (\bibinfo {year} {2007})},\ \Eprint {http://arxiv.org/abs/0704.1615}
  {arXiv:0704.1615 [nucl-th]} \BibitemShut {NoStop}%
\bibitem [{\citenamefont {Arndt}\ \emph {et~al.}(2006)\citenamefont {Arndt},
  \citenamefont {Briscoe}, \citenamefont {Strakovsky},\ and\ \citenamefont
  {Workman}}]{Arndt:2006bf}%
  \BibitemOpen
  \bibfield  {author} {\bibinfo {author} {\bibfnamefont {R.~A.}\ \bibnamefont
  {Arndt}}, \bibinfo {author} {\bibfnamefont {W.~J.}\ \bibnamefont {Briscoe}},
  \bibinfo {author} {\bibfnamefont {I.~I.}\ \bibnamefont {Strakovsky}}, \ and\
  \bibinfo {author} {\bibfnamefont {R.~L.}\ \bibnamefont {Workman}},\ }\href
  {\doibase 10.1103/PhysRevC.74.045205} {\bibfield  {journal} {\bibinfo
  {journal} {Phys. Rev.}\ }\textbf {\bibinfo {volume} {C74}},\ \bibinfo {pages}
  {045205} (\bibinfo {year} {2006})},\ \Eprint
  {http://arxiv.org/abs/nucl-th/0605082} {arXiv:nucl-th/0605082 [nucl-th]}
  \BibitemShut {NoStop}%
\bibitem [{\citenamefont {Ronchen}\ \emph {et~al.}(2013)\citenamefont
  {Ronchen}, \citenamefont {Doring}, \citenamefont {Huang}, \citenamefont
  {Haberzettl}, \citenamefont {Haidenbauer}, \citenamefont {Hanhart},
  \citenamefont {Krewald}, \citenamefont {Meissner},\ and\ \citenamefont
  {Nakayama}}]{Ronchen:2012eg}%
  \BibitemOpen
  \bibfield  {author} {\bibinfo {author} {\bibfnamefont {D.}~\bibnamefont
  {Ronchen}}, \bibinfo {author} {\bibfnamefont {M.}~\bibnamefont {Doring}},
  \bibinfo {author} {\bibfnamefont {F.}~\bibnamefont {Huang}}, \bibinfo
  {author} {\bibfnamefont {H.}~\bibnamefont {Haberzettl}}, \bibinfo {author}
  {\bibfnamefont {J.}~\bibnamefont {Haidenbauer}}, \bibinfo {author}
  {\bibfnamefont {C.}~\bibnamefont {Hanhart}}, \bibinfo {author} {\bibfnamefont
  {S.}~\bibnamefont {Krewald}}, \bibinfo {author} {\bibfnamefont {U.~G.}\
  \bibnamefont {Meissner}}, \ and\ \bibinfo {author} {\bibfnamefont
  {K.}~\bibnamefont {Nakayama}},\ }\href {\doibase 10.1140/epja/i2013-13044-5}
  {\bibfield  {journal} {\bibinfo  {journal} {Eur. Phys. J.}\ }\textbf
  {\bibinfo {volume} {A49}},\ \bibinfo {pages} {44} (\bibinfo {year} {2013})},\
  \Eprint {http://arxiv.org/abs/1211.6998} {arXiv:1211.6998 [nucl-th]}
  \BibitemShut {NoStop}%
\bibitem [{GWd()}]{GWdataAnalysisCentre}%
  \BibitemOpen
  \href@noop {} {\enquote {\bibinfo {title} {{GW Data Analysis Center,
  Institute for Nuclear Studies, The George Washington University}},}\
  }\bibinfo {howpublished}
  {\url{http://gwdac.phys.gwu.edu/analysis/pin_analysis.html}},\ \bibinfo
  {note} {accessed: July 2017}\BibitemShut {NoStop}%
\bibitem [{\citenamefont {Suzuki}\ \emph {et~al.}(2009)\citenamefont {Suzuki},
  \citenamefont {Sato},\ and\ \citenamefont {Lee}}]{Suzuki:2008rp}%
  \BibitemOpen
  \bibfield  {author} {\bibinfo {author} {\bibfnamefont {N.}~\bibnamefont
  {Suzuki}}, \bibinfo {author} {\bibfnamefont {T.}~\bibnamefont {Sato}}, \ and\
  \bibinfo {author} {\bibfnamefont {T.~S.~H.}\ \bibnamefont {Lee}},\ }\href
  {\doibase 10.1103/PhysRevC.79.025205} {\bibfield  {journal} {\bibinfo
  {journal} {Phys. Rev.}\ }\textbf {\bibinfo {volume} {C79}},\ \bibinfo {pages}
  {025205} (\bibinfo {year} {2009})},\ \Eprint {http://arxiv.org/abs/0806.2043}
  {arXiv:0806.2043 [nucl-th]} \BibitemShut {NoStop}%
\bibitem [{\citenamefont {Doring}\ \emph {et~al.}(2009)\citenamefont {Doring},
  \citenamefont {Hanhart}, \citenamefont {Huang}, \citenamefont {Krewald},\
  and\ \citenamefont {Meissner}}]{Doring:2009yv}%
  \BibitemOpen
  \bibfield  {author} {\bibinfo {author} {\bibfnamefont {M.}~\bibnamefont
  {Doring}}, \bibinfo {author} {\bibfnamefont {C.}~\bibnamefont {Hanhart}},
  \bibinfo {author} {\bibfnamefont {F.}~\bibnamefont {Huang}}, \bibinfo
  {author} {\bibfnamefont {S.}~\bibnamefont {Krewald}}, \ and\ \bibinfo
  {author} {\bibfnamefont {U.~G.}\ \bibnamefont {Meissner}},\ }\href {\doibase
  10.1016/j.nuclphysa.2009.08.010} {\bibfield  {journal} {\bibinfo  {journal}
  {Nucl. Phys.}\ }\textbf {\bibinfo {volume} {A829}},\ \bibinfo {pages} {170}
  (\bibinfo {year} {2009})},\ \Eprint {http://arxiv.org/abs/0903.4337}
  {arXiv:0903.4337 [nucl-th]} \BibitemShut {NoStop}%
\bibitem [{\citenamefont {Lang}\ \emph {et~al.}(2017)\citenamefont {Lang},
  \citenamefont {Leskovec}, \citenamefont {Padmanath},\ and\ \citenamefont
  {Prelovsek}}]{Lang:2016hnn}%
  \BibitemOpen
  \bibfield  {author} {\bibinfo {author} {\bibfnamefont {C.~B.}\ \bibnamefont
  {Lang}}, \bibinfo {author} {\bibfnamefont {L.}~\bibnamefont {Leskovec}},
  \bibinfo {author} {\bibfnamefont {M.}~\bibnamefont {Padmanath}}, \ and\
  \bibinfo {author} {\bibfnamefont {S.}~\bibnamefont {Prelovsek}},\ }\href
  {\doibase 10.1103/PhysRevD.95.014510} {\bibfield  {journal} {\bibinfo
  {journal} {Phys. Rev.}\ }\textbf {\bibinfo {volume} {D95}},\ \bibinfo {pages}
  {014510} (\bibinfo {year} {2017})},\ \Eprint
  {http://arxiv.org/abs/1610.01422} {arXiv:1610.01422 [hep-lat]} \BibitemShut
  {NoStop}%
\bibitem [{\citenamefont {Mahbub}\ \emph {et~al.}(2012)\citenamefont {Mahbub},
  \citenamefont {Kamleh}, \citenamefont {Leinweber}, \citenamefont {Moran},\
  and\ \citenamefont {Williams}}]{Mahbub:2010rm}%
  \BibitemOpen
  \bibfield  {author} {\bibinfo {author} {\bibfnamefont {M.~S.}\ \bibnamefont
  {Mahbub}}, \bibinfo {author} {\bibfnamefont {W.}~\bibnamefont {Kamleh}},
  \bibinfo {author} {\bibfnamefont {D.~B.}\ \bibnamefont {Leinweber}}, \bibinfo
  {author} {\bibfnamefont {P.~J.}\ \bibnamefont {Moran}}, \ and\ \bibinfo
  {author} {\bibfnamefont {A.~G.}\ \bibnamefont {Williams}} (\bibinfo
  {collaboration} {CSSM Lattice}),\ }\href {\doibase
  10.1016/j.physletb.2011.12.048} {\bibfield  {journal} {\bibinfo  {journal}
  {Phys. Lett.}\ }\textbf {\bibinfo {volume} {B707}},\ \bibinfo {pages} {389}
  (\bibinfo {year} {2012})},\ \Eprint {http://arxiv.org/abs/1011.5724}
  {arXiv:1011.5724 [hep-lat]} \BibitemShut {NoStop}%
\bibitem [{\citenamefont {Mahbub}\ \emph
  {et~al.}(2013{\natexlab{a}})\citenamefont {Mahbub}, \citenamefont {Kamleh},
  \citenamefont {Leinweber}, \citenamefont {Moran},\ and\ \citenamefont
  {Williams}}]{Mahbub:2012ri}%
  \BibitemOpen
  \bibfield  {author} {\bibinfo {author} {\bibfnamefont {M.~S.}\ \bibnamefont
  {Mahbub}}, \bibinfo {author} {\bibfnamefont {W.}~\bibnamefont {Kamleh}},
  \bibinfo {author} {\bibfnamefont {D.~B.}\ \bibnamefont {Leinweber}}, \bibinfo
  {author} {\bibfnamefont {P.~J.}\ \bibnamefont {Moran}}, \ and\ \bibinfo
  {author} {\bibfnamefont {A.~G.}\ \bibnamefont {Williams}},\ }\href {\doibase
  10.1103/PhysRevD.87.011501} {\bibfield  {journal} {\bibinfo  {journal} {Phys.
  Rev.}\ }\textbf {\bibinfo {volume} {D87}},\ \bibinfo {pages} {011501}
  (\bibinfo {year} {2013}{\natexlab{a}})},\ \Eprint
  {http://arxiv.org/abs/1209.0240} {arXiv:1209.0240 [hep-lat]} \BibitemShut
  {NoStop}%
\bibitem [{\citenamefont {Mahbub}\ \emph
  {et~al.}(2013{\natexlab{b}})\citenamefont {Mahbub}, \citenamefont {Kamleh},
  \citenamefont {Leinweber}, \citenamefont {Moran},\ and\ \citenamefont
  {Williams}}]{Mahbub:2013ala}%
  \BibitemOpen
  \bibfield  {author} {\bibinfo {author} {\bibfnamefont {M.~S.}\ \bibnamefont
  {Mahbub}}, \bibinfo {author} {\bibfnamefont {W.}~\bibnamefont {Kamleh}},
  \bibinfo {author} {\bibfnamefont {D.~B.}\ \bibnamefont {Leinweber}}, \bibinfo
  {author} {\bibfnamefont {P.~J.}\ \bibnamefont {Moran}}, \ and\ \bibinfo
  {author} {\bibfnamefont {A.~G.}\ \bibnamefont {Williams}},\ }\href {\doibase
  10.1103/PhysRevD.87.094506} {\bibfield  {journal} {\bibinfo  {journal} {Phys.
  Rev.}\ }\textbf {\bibinfo {volume} {D87}},\ \bibinfo {pages} {094506}
  (\bibinfo {year} {2013}{\natexlab{b}})},\ \Eprint
  {http://arxiv.org/abs/1302.2987} {arXiv:1302.2987 [hep-lat]} \BibitemShut
  {NoStop}%
\bibitem [{\citenamefont {Mahbub}\ \emph {et~al.}(2014)\citenamefont {Mahbub},
  \citenamefont {Kamleh}, \citenamefont {Leinweber},\ and\ \citenamefont
  {Williams}}]{Mahbub:2013bba}%
  \BibitemOpen
  \bibfield  {author} {\bibinfo {author} {\bibfnamefont {M.~S.}\ \bibnamefont
  {Mahbub}}, \bibinfo {author} {\bibfnamefont {W.}~\bibnamefont {Kamleh}},
  \bibinfo {author} {\bibfnamefont {D.~B.}\ \bibnamefont {Leinweber}}, \ and\
  \bibinfo {author} {\bibfnamefont {A.~G.}\ \bibnamefont {Williams}},\ }\href
  {\doibase 10.1016/j.aop.2014.01.004} {\bibfield  {journal} {\bibinfo
  {journal} {Annals Phys.}\ }\textbf {\bibinfo {volume} {342}},\ \bibinfo
  {pages} {270} (\bibinfo {year} {2014})},\ \Eprint
  {http://arxiv.org/abs/1310.6803} {arXiv:1310.6803 [hep-lat]} \BibitemShut
  {NoStop}%
\bibitem [{\citenamefont {Kiratidis}\ \emph {et~al.}(2015)\citenamefont
  {Kiratidis}, \citenamefont {Kamleh}, \citenamefont {Leinweber},\ and\
  \citenamefont {Owen}}]{Kiratidis:2015vpa}%
  \BibitemOpen
  \bibfield  {author} {\bibinfo {author} {\bibfnamefont {A.~L.}\ \bibnamefont
  {Kiratidis}}, \bibinfo {author} {\bibfnamefont {W.}~\bibnamefont {Kamleh}},
  \bibinfo {author} {\bibfnamefont {D.~B.}\ \bibnamefont {Leinweber}}, \ and\
  \bibinfo {author} {\bibfnamefont {B.~J.}\ \bibnamefont {Owen}},\ }\href
  {\doibase 10.1103/PhysRevD.91.094509} {\bibfield  {journal} {\bibinfo
  {journal} {Phys. Rev.}\ }\textbf {\bibinfo {volume} {D91}},\ \bibinfo {pages}
  {094509} (\bibinfo {year} {2015})},\ \Eprint
  {http://arxiv.org/abs/1501.07667} {arXiv:1501.07667 [hep-lat]} \BibitemShut
  {NoStop}%
\bibitem [{\citenamefont {Kiratidis}\ \emph {et~al.}(2017)\citenamefont
  {Kiratidis}, \citenamefont {Kamleh}, \citenamefont {Leinweber}, \citenamefont
  {Liu}, \citenamefont {Stokes},\ and\ \citenamefont
  {Thomas}}]{Kiratidis:2016hda}%
  \BibitemOpen
  \bibfield  {author} {\bibinfo {author} {\bibfnamefont {A.~L.}\ \bibnamefont
  {Kiratidis}}, \bibinfo {author} {\bibfnamefont {W.}~\bibnamefont {Kamleh}},
  \bibinfo {author} {\bibfnamefont {D.~B.}\ \bibnamefont {Leinweber}}, \bibinfo
  {author} {\bibfnamefont {Z.-W.}\ \bibnamefont {Liu}}, \bibinfo {author}
  {\bibfnamefont {F.~M.}\ \bibnamefont {Stokes}}, \ and\ \bibinfo {author}
  {\bibfnamefont {A.~W.}\ \bibnamefont {Thomas}},\ }\href {\doibase
  10.1103/PhysRevD.95.074507} {\bibfield  {journal} {\bibinfo  {journal} {Phys.
  Rev.}\ }\textbf {\bibinfo {volume} {D95}},\ \bibinfo {pages} {074507}
  (\bibinfo {year} {2017})},\ \Eprint {http://arxiv.org/abs/1608.03051}
  {arXiv:1608.03051 [hep-lat]} \BibitemShut {NoStop}%
\bibitem [{\citenamefont {Dudek}\ and\ \citenamefont
  {Edwards}(2012)}]{Dudek:2012ag}%
  \BibitemOpen
  \bibfield  {author} {\bibinfo {author} {\bibfnamefont {J.~J.}\ \bibnamefont
  {Dudek}}\ and\ \bibinfo {author} {\bibfnamefont {R.~G.}\ \bibnamefont
  {Edwards}},\ }\href {\doibase 10.1103/PhysRevD.85.054016} {\bibfield
  {journal} {\bibinfo  {journal} {Phys. Rev.}\ }\textbf {\bibinfo {volume}
  {D85}},\ \bibinfo {pages} {054016} (\bibinfo {year} {2012})},\ \Eprint
  {http://arxiv.org/abs/1201.2349} {arXiv:1201.2349 [hep-ph]} \BibitemShut
  {NoStop}%
\bibitem [{\citenamefont {{B\"ar}}(2017)}]{Bar:2017kxh}%
  \BibitemOpen
  \bibfield  {author} {\bibinfo {author} {\bibfnamefont {O.}~\bibnamefont
  {{B\"ar}}},\ }\href {\doibase 10.1142/S0217751X17300113} {\bibfield
  {journal} {\bibinfo  {journal} {Int. J. Mod. Phys.}\ }\textbf {\bibinfo
  {volume} {A32}},\ \bibinfo {pages} {1730011} (\bibinfo {year} {2017})},\
  \Eprint {http://arxiv.org/abs/1705.02806} {arXiv:1705.02806 [hep-lat]}
  \BibitemShut {NoStop}%
\bibitem [{\citenamefont {Alexandrou}\ \emph {et~al.}(2015)\citenamefont
  {Alexandrou}, \citenamefont {Leontiou}, \citenamefont {Papanicolas},\ and\
  \citenamefont {Stiliaris}}]{Alexandrou:2014mka}%
  \BibitemOpen
  \bibfield  {author} {\bibinfo {author} {\bibfnamefont {C.}~\bibnamefont
  {Alexandrou}}, \bibinfo {author} {\bibfnamefont {T.}~\bibnamefont
  {Leontiou}}, \bibinfo {author} {\bibfnamefont {C.~N.}\ \bibnamefont
  {Papanicolas}}, \ and\ \bibinfo {author} {\bibfnamefont {E.}~\bibnamefont
  {Stiliaris}},\ }\href {\doibase 10.1103/PhysRevD.91.014506} {\bibfield
  {journal} {\bibinfo  {journal} {Phys. Rev.}\ }\textbf {\bibinfo {volume}
  {D91}},\ \bibinfo {pages} {014506} (\bibinfo {year} {2015})},\ \Eprint
  {http://arxiv.org/abs/1411.6765} {arXiv:1411.6765 [hep-lat]} \BibitemShut
  {NoStop}%
\bibitem [{\citenamefont {Roberts}\ \emph {et~al.}(2013)\citenamefont
  {Roberts}, \citenamefont {Kamleh},\ and\ \citenamefont
  {Leinweber}}]{Roberts:2013ipa}%
  \BibitemOpen
  \bibfield  {author} {\bibinfo {author} {\bibfnamefont {D.~S.}\ \bibnamefont
  {Roberts}}, \bibinfo {author} {\bibfnamefont {W.}~\bibnamefont {Kamleh}}, \
  and\ \bibinfo {author} {\bibfnamefont {D.~B.}\ \bibnamefont {Leinweber}},\
  }\href {\doibase 10.1016/j.physletb.2013.06.056} {\bibfield  {journal}
  {\bibinfo  {journal} {Phys. Lett.}\ }\textbf {\bibinfo {volume} {B725}},\
  \bibinfo {pages} {164} (\bibinfo {year} {2013})},\ \Eprint
  {http://arxiv.org/abs/1304.0325} {arXiv:1304.0325 [hep-lat]} \BibitemShut
  {NoStop}%
\bibitem [{\citenamefont {Roberts}\ \emph {et~al.}(2014)\citenamefont
  {Roberts}, \citenamefont {Kamleh},\ and\ \citenamefont
  {Leinweber}}]{Roberts:2013oea}%
  \BibitemOpen
  \bibfield  {author} {\bibinfo {author} {\bibfnamefont {D.~S.}\ \bibnamefont
  {Roberts}}, \bibinfo {author} {\bibfnamefont {W.}~\bibnamefont {Kamleh}}, \
  and\ \bibinfo {author} {\bibfnamefont {D.~B.}\ \bibnamefont {Leinweber}},\
  }\href {\doibase 10.1103/PhysRevD.89.074501} {\bibfield  {journal} {\bibinfo
  {journal} {Phys. Rev.}\ }\textbf {\bibinfo {volume} {D89}},\ \bibinfo {pages}
  {074501} (\bibinfo {year} {2014})},\ \Eprint {http://arxiv.org/abs/1311.6626}
  {arXiv:1311.6626 [hep-lat]} \BibitemShut {NoStop}%
\bibitem [{\citenamefont {Hansen}\ and\ \citenamefont
  {Sharpe}(2014)}]{Hansen:2014eka}%
  \BibitemOpen
  \bibfield  {author} {\bibinfo {author} {\bibfnamefont {M.~T.}\ \bibnamefont
  {Hansen}}\ and\ \bibinfo {author} {\bibfnamefont {S.~R.}\ \bibnamefont
  {Sharpe}},\ }\href {\doibase 10.1103/PhysRevD.90.116003} {\bibfield
  {journal} {\bibinfo  {journal} {Phys. Rev.}\ }\textbf {\bibinfo {volume}
  {D90}},\ \bibinfo {pages} {116003} (\bibinfo {year} {2014})},\ \Eprint
  {http://arxiv.org/abs/1408.5933} {arXiv:1408.5933 [hep-lat]} \BibitemShut
  {NoStop}%
\bibitem [{\citenamefont {Hansen}\ and\ \citenamefont
  {Sharpe}(2015)}]{Hansen:2015zga}%
  \BibitemOpen
  \bibfield  {author} {\bibinfo {author} {\bibfnamefont {M.~T.}\ \bibnamefont
  {Hansen}}\ and\ \bibinfo {author} {\bibfnamefont {S.~R.}\ \bibnamefont
  {Sharpe}},\ }\href {\doibase 10.1103/PhysRevD.92.114509} {\bibfield
  {journal} {\bibinfo  {journal} {Phys. Rev.}\ }\textbf {\bibinfo {volume}
  {D92}},\ \bibinfo {pages} {114509} (\bibinfo {year} {2015})},\ \Eprint
  {http://arxiv.org/abs/1504.04248} {arXiv:1504.04248 [hep-lat]} \BibitemShut
  {NoStop}%
\end{thebibliography}%

%
%
%

%


\end{document}